\documentclass[sigconf]{acmart}

\usepackage{tikz}
\usepackage{makecell}
\usepackage{siunitx}
\usepackage[most]{tcolorbox} 
\usepackage{hyperref}        
\usepackage[nameinlink,noabbrev]{cleveref}
\usepackage{xparse}

\newcommand{\circled}[1]{\tikz[baseline=(char.base)]{
  \node[shape=circle,draw,inner sep=1pt] (char) {#1};}}

\usepackage{listings}
\usepackage{subcaption}
\usepackage[most]{tcolorbox}
\Crefname{lstlisting}{Listing}{Listings}

\AtBeginDocument{%
  }

\renewcommand\footnotetextcopyrightpermission[1]{}

\setcopyright{None}
\copyrightyear{}
\acmYear{}
\acmConference[xxx]{xxx}{2026}{xxx}
\usepackage[all=normal,paragraphs=normal,wordspacing=normal,charwidths=normal,indent=normal,floats=normal,leading=normal,lists=normal,bibnotes=normal]{savetrees}

\begin{document}

\newcounter{finding}
\renewcommand{\thefinding}{\arabic{finding}}

\newtcolorbox{findingbanner}[1][]{
  enhanced, breakable, sharp corners,
  boxrule=0.6pt,
  left=6pt,right=6pt,top=5pt,bottom=5pt,
  colback=white, colframe=orange,
  fonttitle=\bfseries,
  title={Finding~\thefinding},
  before upper={\refstepcounter{finding}},
  #1
}

\newcommand{\findingref}[1]{\hyperref[#1]{Finding~\ref*{#1}}}

\title{Revealing NVIDIA Closed-Source Driver Command Streams for CPU–GPU Runtime Behavior Insight}


\author{Yuang Yan}
\email{yuang.yan@queensu.ca}
\orcid{0009-0009-2289-5945}
\affiliation{%
  \institution{Queen's University}
  \city{Kingston}
  \state{Ontario}
  \country{Canada}
}
\author{Ian Karlin}
\email{ian.karlin@queensu.ca}
\affiliation{%
  \institution{Queen's University}
  \city{Kingston}
  \state{Ontario}
  \country{Canada}
}
\author{Ryan Grant}
\email{ryan.grant@queensu.ca}
\affiliation{%
  \institution{Queen's University}
  \city{Kingston}
  \state{Ontario}
  \country{Canada}
}

\begin{abstract}

For NVIDIA GPUs, CUDA is the primary interface through which applications orchestrate GPU execution, yet much of the logic that realizes CUDA operations resides in NVIDIA’s closed-source userspace driver. As a result, the translation from high-level CUDA APIs to low-level hardware commands remains opaque, limiting both software understanding and performance attribution.

This paper makes that command path visible. We recover the hardware command streams emitted by NVIDIA’s closed-source userspace driver with full integrity by leveraging the recently open-sourced kernel driver, instrumenting the memory-mapping path, and installing a hardware watchpoint on the userspace mapping of the GPU doorbell register. This lets us capture complete command submissions at the moment they are committed.

Using this methodology, we present two case studies. For CUDA data movement, we identify the DMA submission modes selected by the driver and characterize their raw hardware performance independently of driver overhead through CUDA-bypassing controlled command issuance. For CUDA Graphs, we show that the reduced launch overhead in newer CUDA releases is associated with a smaller command footprint and a more efficient submission pattern. Together, these results show that command-level visibility provides a practical basis for understanding and optimizing GPU middleware behavior, improving performance interpretation, and informing future hardware--software co-design for CUDA and related accelerator stacks.



\end{abstract}




\maketitle

\section{Introduction}

GPUs have become indispensable for scientific computing and AI due to their massive parallel compute capability. In modern computing systems, CPUs and GPUs are separate processors with distinct roles: the CPU remains the control center and the primary interface to the operating system, while the GPU executes offloaded computational work. As CPUs and GPUs are distinct processing units connected through the system interconnect, their execution is not intrinsically unified.

To coordinate CPUs and GPUs efficiently, one representative software framework is NVIDIA's CUDA, which provides a programming interface for issuing work to the GPU from ordinary host programs. In the CUDA model, code executed on the GPU (i.e., kernels) can run concurrently with code on the CPU, thereby exploiting system heterogeneity. In addition, the GPU can overlap host--device data transfer with its own computation. CUDA abstracts different parallelisms through \textit{streams}: operations issued to different streams are independent, while operations within the same stream follow program order. CUDA also provides interfaces for launching kernels, initiating data transfers, and synchronizing device progress so that the host can determine when GPU work has completed. In modern scientific applications, CUDA becomes the default way to orchestrate GPU execution, with no more direct interface exposed to users.

\begin{figure}[h]
  \centering
  \includegraphics[width=0.75\columnwidth]{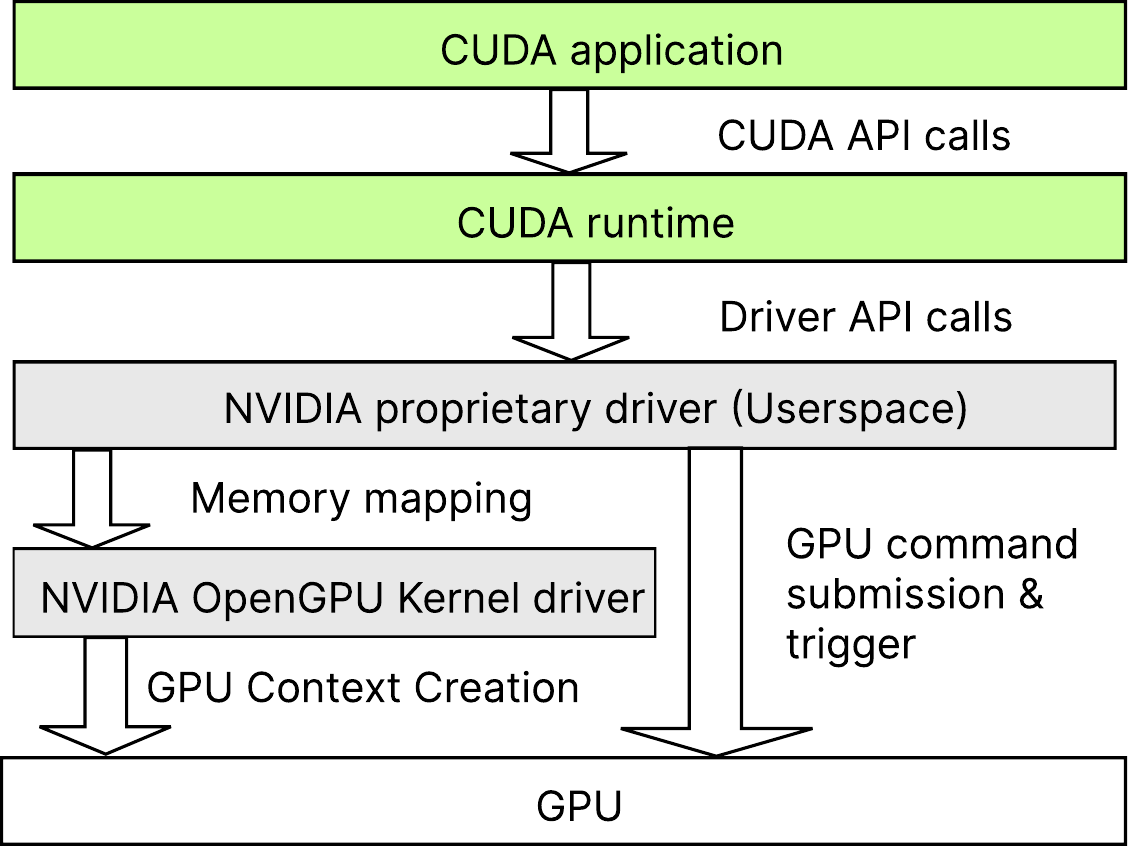}
  \caption{CUDA software stack with layer responsibilities and key interaction paths to the GPU}
  \label{fig:stack}
\end{figure}

The functionality of CUDA APIs, however is not provided by the CUDA runtime alone, but largely by the GPU driver underneath it, which serves as the middleware layer between host software and GPU hardware. \Cref{fig:stack} shows the hierarchy of this supporting software stack. When a CUDA routine involves GPU work, the driver translates the high-level API request into hardware commands and submits them to the device. The driver consists of two parts: a \textit{userspace} driver and a \textit{kernel} driver. Most requests are handled in userspace, while operations that require operating-system privileges, such as memory mapping, are forwarded to the kernel driver through system calls.

Because this driver layer mediates communication between the host and the GPU, application performance depends not only on the hardware capability itself, but also on how the driver interacts with the GPU and on the overhead introduced by the driver’s own processing. This middleware overhead has also become increasingly important as GPU computation grows more optimized and intense, making software-side overheads harder to ignore. Techniques such as kernel fusion~\cite{kernelfusion} and CUDA Graphs~\cite{cudagraph} have therefore been introduced, in part, to reduce this cost.

Despite the important role of the GPU driver in efficient host--device interaction, analyzing it remains technically difficult because much of the driver stack remains difficult to inspect directly (grey layers in~\Cref{fig:stack}). Although NVIDIA has recently open-sourced the kernel driver~\cite{opengpu}, most CUDA operation-handling logic still resides in the userspace driver, which remains proprietary.


As a result, the mapping from high-level software interfaces to low-level hardware commands remains opaque. Without such knowledge, it is difficult to understand the established mechanisms of CPU--GPU interaction and to extract design lessons from them. This opacity also limits broader community efforts to reason about, improve, or adapt similar accelerator software stacks, and hinders collaborative optimization of CUDA’s internal driver behavior for the benefit of applications built on top of it.

A second implication is performance interpretation. Tools such as Nsight Systems~\cite{nvidia-nsight-systems} report timings at the runtime level, where costs from multiple stages along the software--hardware path are aggregated into a single duration. As a result, the contribution of each stage is not directly visible, making it difficult to isolate raw device behavior, especially in non-vendor-built environments such as disaggregated systems, where software-side costs can dominate small-message performance.

A deeper understanding of device behavior motivates this work. The goal is: to recover the hardware command streams generated from CUDA API calls by the closed-source userspace driver, and to enable CUDA-bypassing hardware control for direct measurement of raw data-transfer performance. We achieve this by analyzing the command-submission mechanism in detail, as presented in~\Cref{sec:subarch}, and by instrumenting the kernel-driver memory-mapping path together with CPU debug registers, with high level overview in~\Cref{sec:techchallenges} and details in~\Cref{sec:method}. Specifically, our contributions are as follows:

\subsection*{Contributions}

\begin{itemize}
    \item \textbf{Command-stream capture and reconstruction}
    We capture, reconstruct, and parse the full runtime command stream emitted by NVIDIA’s closed-source user-space driver, (\Cref{lst:decoded}). By examining the translated hardware commands, we gain insight into how high-level CUDA abstractions are implemented at the hardware level, including mechanisms such as stream ordering and event recording.

    \item \textbf{Data-Movement Analysis and Controlled DMA Launch}
    We analyze CUDA data-movement behavior, revealing how the driver switches between DMA submission modes which employs different driver path as transfer size varies. Beyond observation, we introduce a CUDA-bypassing mechanism that directly programs the DMA engines to perform the same data movement as \texttt{cudaMemcpy}, enabling controlled measurements that decouple raw GPU DMA-engine performance from driver overhead.

   \item \textbf{Lesson for reducing driver overhead from CUDA Graph}
    We analyze CUDA Graph execution, a mechanism for reducing repeated CUDA runtime launch overhead, and show how newer CUDA releases lower host-side launch cost by reshaping the command stream and submission pattern, thereby reducing CPU involvement in the critical path.
\end{itemize}

Together, the two case studies (data movement and CUDA Graphs) demonstrate the benefit of making the driver’s translation into hardware commands visible. They represent only the first examples explored in this paper, not the limit of the methodology. The same approach can be applied to other CUDA APIs and applications of interest. More generally, the visibility provided by this broadly applicable method helps the community better understand and improve accelerator software stacks.

\section{Related Work}
Since NVIDIA open-sourced its GPU kernel driver in 2022, a growing body of work has begun to analyze this operating-system-level software stack from several perspectives. Existing studies have examined GPU context scheduling and management~\cite{gpuinternals}, virtualized GPU sharing for cloud services~\cite{kernelinspec}, and the behavior of CUDA Unified Virtual Memory (UVM)~\cite{uvm}. A substantial portion of this literature is security-oriented. This includes studies of NVIDIA Confidential Computing~\cite{ccomputing}, systems for confidential collaborative learning and privacy-preserving GPU services~\cite{gputravel,XpuTEE}, forensic analysis of driver state for detecting malicious GPU activity~\cite{gpuforenisc}, and side-channel attacks on MIG through reverse engineering of GPU TLBs~\cite{TunneLs}.

Some of these works modify the driver or discuss the structure of GPU command submission and related data paths at a conceptual level. For example, Allen et al.~\cite{uvm} instrument the driver to collect fine-grained fault metadata, while Zhang et al.~\cite{kernelinspec} intercept GPU command buffers in kernel space to support multi-tenant GPU sharing and isolation. However, these works do not reconstruct the end-to-end command trajectory of high-level CUDA API calls as emitted by NVIDIA's closed-source user-space driver, as done in this work.





\section{Technical Challenges and Our Approach}
\label{sec:techchallenges}
For performance, the closed-source user-space driver maps the GPU's PCIe BAR (Base Address Register) region into its virtual address space, allowing it to issue command and doorbell writes directly to GPU registers without entering the kernel. This design avoids frequent user–kernel context switches and makes GPU notification more immediate. \cite{microsoftdb} However this makes the submission path difficult to observe.

A natural approach is to periodically poll the memory regions established by the open-source kernel driver. One prior effort \cite{gpuinternals} sampled MMIO (Memory-Mapped IO) execution registers to study GPU context preemption. While polling can work for byte-wide registers, it is inadequate for capturing the much larger command stream. Its limited sampling rate cannot guarantee that every submission is observed, given the command stream size and the high frequency of GPU updates. Moreover, without intervening in the submission path, sampled register and memory state may be inconsistent, producing partial or corrupted views of in-flight commands.

We configure the CPU debug registers to watch the userspace virtual address mapped to the GPU doorbell, an MMIO register used to notify the GPU of new work. When this write occurs, the watchpoint traps into the kernel and pauses the userspace driver, creating a static window in which we can inspect the command buffer and recover a complete, intact view of the newly submitted command stream.

This mechanism creates an observation window at the submission boundary. By pausing the process when a submission is committed, we capture each submission cycle as a complete, intact unit instead of a partial or fragmented view.



\section{NVIDIA GPU Command Submission Architecture}
\label{sec:subarch}

This section provides an overview of NVIDIA command submission based on the OpenGPU Kernel Modules source code~\cite{nvidia_open_gpu_kernel_modules_580_105_08} and its documentation~\cite{nvidia_open_gpu_doc}. Unless otherwise noted, the descriptions in this section are derived from these sources. We first explain how work is packaged into command buffers and queued for execution. Then we introduce the command encoding needed to interpret parsed streams. We next explain how command submission and progress state are tracked and synchronized between driver data structures and hardware registers, which later supports command reconstruction (\Cref{subsec:reconstruction}). Finally, we discuss how CUDA synchronization and timing map onto GPU command primitives, enabling device-side timing of custom-issued operations in \Cref{subsec:customizecmd}.

\subsection{Command Submission Hierarchy}
\label{sec:submission_hierarchy}
\begin{figure}[h]
  \centering
  \includegraphics[width=\linewidth]{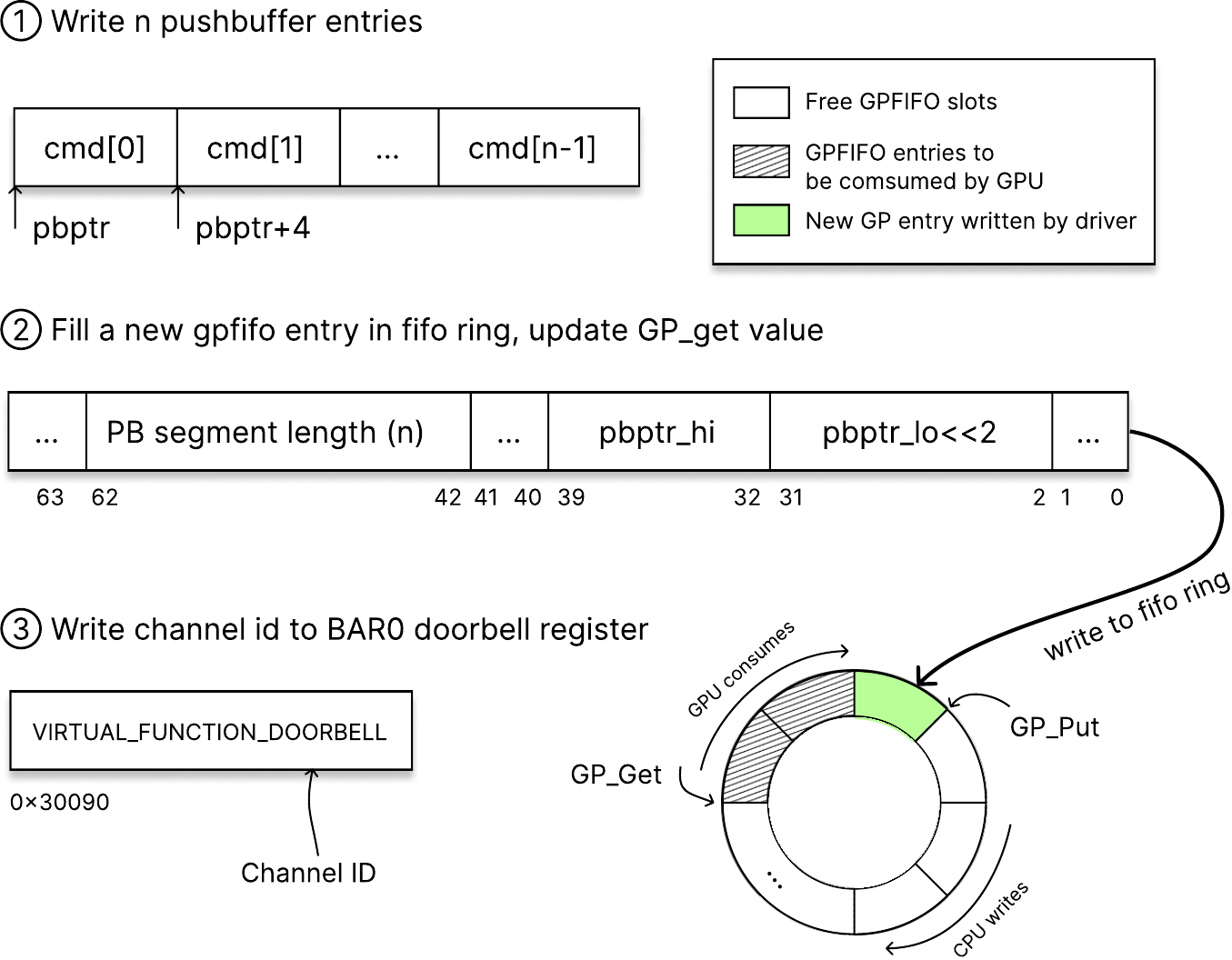}
  \caption{Driver command submission path: pushbuffer writes, GPFIFO enqueue, and doorbell notification.}
  \label{fig:cmdsubmission}
\end{figure}

NVIDIA GPUs employ a two-level command submission hierarchy composed of the
\textbf{GPFIFO} (Get/Put FIFO) and the \textbf{pushbuffer} (PB). A typical command
submission path is depicted in \Cref{fig:cmdsubmission}.

The pushbuffer holds the raw command stream that is directly consumed by GPU
engines. Each PB entry is 4 bytes. To submit a batch of work, the
driver first writes the translated commands from higher-level API calls (e.g.,
encoding a \texttt{cudaMemcpy()} into copy/DMA descriptors) into the pushbuffer
(\circled{1}). Next, the driver enqueues a \textbf{GPFIFO entry} (\circled{2}), which encodes the
starting GPU virtual address of the pushbuffer segment and its length 
into a 64-bit descriptor.

The GPFIFO itself is a ring buffer established when a GPU context (channel) is
created (details in \Cref{sec:channel}). The driver acts as the producer,
advancing the producer index \texttt{GP\_Put}, while the GPU consumes entries by
advancing \texttt{GP\_Get}. After filling a new slot, the driver increments
\texttt{GP\_Put} and rings the doorbell (\circled{3}) by writing the channel/context
identifier to a doorbell register. This doorbell register is global across
channels and is accessed via the PCIe BAR0 MMIO range at the
\texttt{VIRTUAL\_FUNCTION\_DOORBELL} offset.

Upon receiving the doorbell notification, the GPU \textbf{PBDMA} (pushbuffer DMA)
engine fetches the newly enqueued GPFIFO entry, reconstructs the pushbuffer
address and length, and then reads the corresponding command stream from the
pushbuffer. The PBDMA front-end parses the stream and dispatches work to the
appropriate engines (e.g., compute, copy), acting as the consumer-side
mirror of the driver’s submission logic.

\subsection{Channel Context Structure}
\label{sec:channel}

In \Cref{sec:submission_hierarchy}, we described how commands are queued and consumed using \texttt{GP\_PUT} and \texttt{GP\_GET}. These pointers, analogous to a CPU thread’s program counter, represents the execution state of a runnable context. A runable context is referred to as a \textbf{Channel}. In addition, a channel contains the memory state that resolves GPU virtual to physical memory addressing, and engine state that specifies the status of the engine executing the work.

 The persistent per-channel state is stored in \textbf{RAMIN} (channel instance memory). The execution state (also known as \emph{host state}) resides in \textbf{RAMFC} (FIFO context memory). During a context switch, the GPU front-end saves and restores per-channel fields between RAMIN and the PBDMA registers, similar to a CPU saving/restoring thread state from the \textit{Program Control Block} (Linux \texttt{task\_struct}).






The RAMFC containing the \texttt{GP\_Put} value is allocated in a privileged region that the
user-mode driver cannot directly access. To support the direct user-space command
submission, NVIDIA exposes a user-accessible   
memory region called \textbf{USERD}. USERD allows the user-mode driver to update
 producer index \texttt{GP\_PUT} through its virtual address mapping.

When the context is in execution, USERD therefore holds the freshest \texttt{GP\_PUT} value
written by the userspace driver. After the driver rings the doorbell, the \textbf{PBDMA}
engine loads the corresponding state from USERD into its execution registers to begin
(or continue) consuming the GPFIFO. When configured, the GPU periodically
writes the consumer index \texttt{GP\_GET} back to USERD. In contrast, the copies of \texttt{GP\_PUT}/\texttt{GP\_GET} stored in
\textbf{RAMFC} remain unchanged while the channel is running, and are only updated when
a context switch saves/restores channel state.

\begin{figure}[h]
  \centering
  \includegraphics[width=\linewidth]{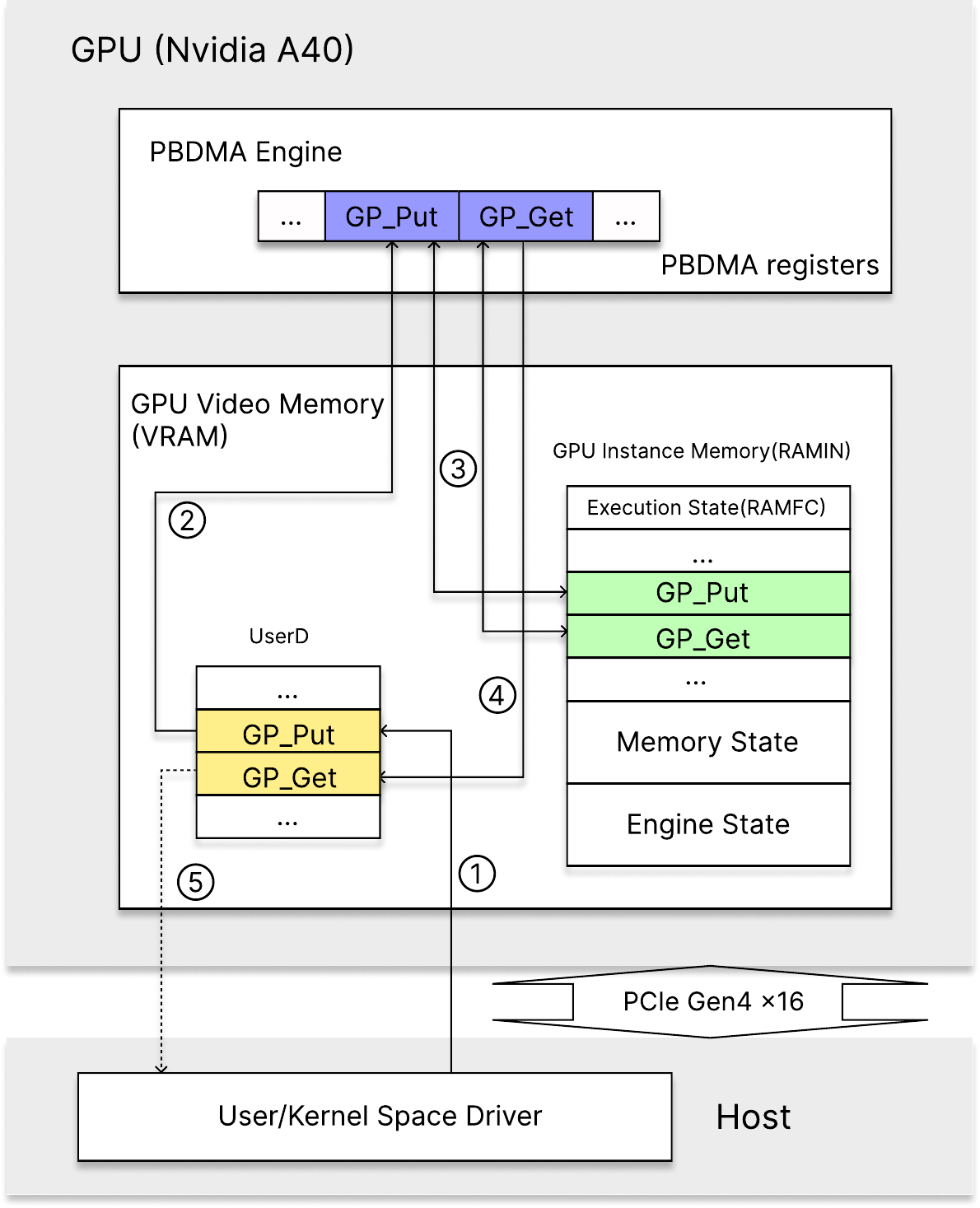}
  \caption{Synchronization of different copies of GPFIFO values. Arrows show how \texttt{GP\_PUT}/\texttt{GP\_GET} values propagate between USERD (userspace window), RAMFC/RAMIN (saved context), and live PBDMA execution registers.}
  \label{fig:pbdma}
\end{figure}

\Cref{fig:pbdma} illustrates how these replicated \texttt{GP\_PUT}/\texttt{GP\_GET}
values are synchronized across USERD, RAMFC, and live PBDMA registers.

\begin{itemize}
  \item \circled{1} The driver writes a pushbuffer segment, assembles a GPFIFO entry,
  and advances \texttt{GP\_PUT} in USERD.
  \item \circled{2} After the doorbell write, PBDMA fetches the latest \texttt{GP\_PUT}
  from USERD into its execution registers.
  \item \circled{3} On a context switch, \texttt{GP\_PUT}/\texttt{GP\_GET} are saved to
  RAMFC when the channel is switched out, and restored back into PBDMA registers when
  the channel is switched in.
  \item \circled{4} If write-back is enabled, the GPU periodically writes
  \texttt{GP\_GET} (the consumer index) back to USERD.
  \item \circled{5} The driver may poll \texttt{GP\_GET} in USERD to track progress,
  However the documentation discourage this manner ~\cite{nvidia_open_gpu_doc}.
\end{itemize}

\subsection{Synchronization and Timing Mechanism}
\label{subsec:timestamp}
In CUDA, operations issued to the same stream are executed in order: a later operation in that stream does not begin until all earlier operations have completed. If subsequent CPU work depends on the completion of previously issued GPU operations, the programmer must explicitly wait using stream synchronization or event synchronization. Stream synchronization blocks until all earlier work in the stream finishes, while event synchronization waits until execution reaches a specific recorded point in the stream, so that dependent CPU code can safely proceed without data hazards from incomplete GPU work~\cite{nvidia_cuda_async_execution}.

Both intra-device and device–host synchronization are implemented using a hardware primitive referred to a \textit{memory semaphore}, which acts as a completion barrier. The driver appends a \textit{semaphore release} command at the end of a sequence of submitted hardware commands (aka. a \textbf{progress tracker} in NVIDIA terms). A semaphore release specifies (i) a target address and (ii) a payload value to be written to that address. Because the engines execute commands in order, the payload appearing at the target address implies that all preceding commands in that submission sequence have completed. Synchronization waits for the semaphore value to match the expected payload, enforcing dependencies in the same way.

The GPU can also be configured to write a timestamp with nanosecond resolution next to the payload location, indicating when the semaphore update occurred. By subtracting the timestamps associated with two semaphore releases, we obtain the elapsed time between the corresponding completion points, which is equivalent to how cudaEventElapsedTime works.

\label{sec:sync}

\section{Command Stream Extraction Methodology}
\label{sec:method}
To restore the trajectory of GPU commands, we trace the submission path in reverse: From the doorbell write through the corresponding GPFIFO entries to the originating pushbuffer, since the doorbell write is the driver’s final commit point.

In \Cref{subsec:getdb}, we describe how we identify the submitting channel when the doorbell is being written. We then present the procedure for recovering the channel’s submission state and extracting the newly enqueued commands in~\Cref{subsec:reconstruction}. Finally, \Cref{subsec:customizecmd} shows how we actively emit customized commands to exert specific mechanisms.

\subsection{Intercepting doorbell writes}
\label{subsec:getdb}

While user space can issue doorbell writes directly, bypassing the kernel for submission as introduced in~\Cref{sec:techchallenges}, it must still rely on the kernel driver to establish the memory mapping to the physical I/O address of the doorbell register. In NVIDIA’s current kernel driver, all such mapping requests are handled by \texttt{nv\_mmap}, where user space supplies the target physical address through the \texttt{mmap} system call.

\begin{figure}[h]
  \centering
  \includegraphics[width=\linewidth]{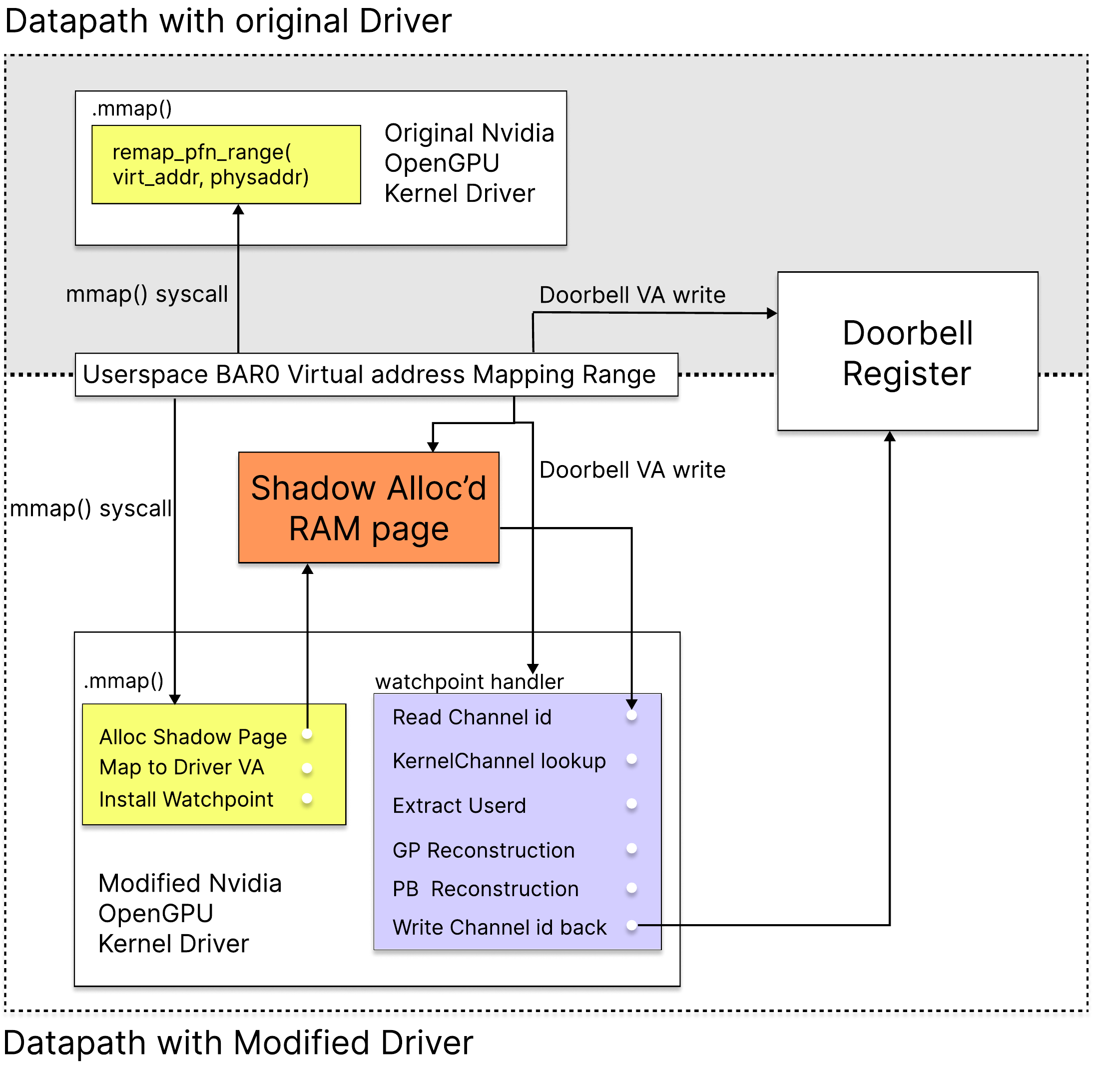}
  \caption{Comparison of mapping and submission paths. Top: In the original driver. Bottom: Our modified driver.}
  \label{fig:method}
\end{figure}

\Cref{fig:method}, shows how the driver works before and after our modifications. The top half (grey block) shows
the data path of doorbell triggers for the original opengpu kernel driver,
where the signal is directly delivered to the mapped doorbell register on GPU.  The bottom half (white block) shows
our modified driver that leverages the mandatory
\texttt{nv\_mmap} step to intercept the user-mode submission path and make doorbell-triggered
submissions observable.

Inside \texttt{nv\_mmap}, if the requested range covers the doorbell register, we install a hardware watchpoint on the corresponding user-space virtual address, similar to a watchpoint set in GDB. When the userspace driver writes to this mapped doorbell address, the watchpoint triggers and execution traps into the kernel, and we use this window to observe memory state and capture the pushbuffer state at the moment of submission in the watchpoint handler callback. Compared with the fault-triggering method, watchpoint installation guarantees that when the callback occurs, the channel ID has already been written. It also keeps execution in kernel space until memory observation is finished, so no new commands can be written during this window. As a result, we can observe a static, integrity-preserving view of the GPFIFO and pushbuffer state. 

When we first implemented this mechanism, the doorbell writes were successfully intercepted. However, reading the doorbell register back from the GPU always returned zero. This behavior suggests that the register is either non-readable or that its contents are flushed immediately after a write. To address this, we allocate a RAM page during the mapping process and use it as a shadow doorbell page. When the user-space driver performs a write, the value first lands in this shadow page. After we perform the memory observation inside the watchpoint handler, we then forward the captured value to the real doorbell register to allow the submission flow to proceed normally.


\subsection{Reconstructing the execution state}
\label{subsec:reconstruction}


When we intercept a doorbell write, the only information we directly observe is the channel identifier. To reconstruct the submitted command stream, we must recover the channel’s backing execution state and the memory regions it references.

In the OpenGPU kernel driver, the \texttt{KernelChannel} structure records memory descriptors for \texttt{USERD}, \texttt{RAMIN}, and \texttt{RAMFC}, which provide the physical addresses of the corresponding regions. We use the intercepted channel ID to locate the channel's \texttt{KernelChannel} object, and then map \texttt{USERD} and \texttt{RAMFC} into the CPU virtual address space so their fields can be read during reconstruction.

Next, we locate the newly enqueued GPFIFO entry, corresponding to \circled{2} in \Cref{fig:cmdsubmission}. We retrieve the \texttt{GP\_PUT} index from \texttt{USERD}, which contains the latest value updated by the driver, and obtain the GPFIFO base address \texttt{GP\_BASE} from \texttt{RAMFC}. The GPU virtual address of the new GPFIFO entry is then computed as:
\[
\texttt{GP\_PUT\_VA} =
\texttt{GP\_BASE} + (\texttt{GP\_PUT} - 1)\times \texttt{GP\_ENTRY\_SIZE},
\]
where \texttt{GP\_ENTRY\_SIZE} is 8 bytes.

After obtaining the GPU virtual address, we resolve the physical address by walking the GPU MMU pagetable We then map the GPFIFO entry’s physical address to
read its contents. From these GPFIFO entries, we extract the pushbuffer segment’s
GPU virtual address and repeat the same translation and mapping steps to read the
pushbuffer instructions. Finally, we parse the pushbuffer commands according to the method format provided by the opensourced driver headers. As an example, \Cref{lst:decoded} shows the reconstructed GPU
command stream generated by a \texttt{cudaMemcpyAsync} transferring 64MB.

\begin{findingbanner}[title={Finding 1: NVIDIA UVM and its implication on driver addressing}]
\label{finding:GPU_VA}
In CUDA, Unified Virtual Memory (UVM) allows a pointer to be dereferenced by both the host and the device (e.g., within CUDA kernels). Under the hood, the UVM kernel module manages a unified virtual address space by tracking mappings and triggering page migration on demand when an access occurs to a page that is not resident on the accessing processor~\cite{uvm}. Recent integration with Linux HMM~\cite{linux_hmm_doc} (heterogeneous memory management) further extends UVM beyond CUDA-managed allocations by leveraging OS support for mirroring page-table entries~\cite{hmm}.

With the adoption of UVM, GPU virtual addresses used in pushbuffer commands are shared with the process's user-space virtual addresses space. In other words, UVM address unification is leveraged internally by the driver: it can emit addresses directly using CPU virtual addresses without an additional GPU-VA translation step, relying on the UVM machinery to maintain consistency of the shared address space. This feature also facilitate our capability to issue a customized command stream to the GPU.
\end{findingbanner}

\begin{lstlisting} [caption={Example debug trace captured at a doorbell interception, including GPFIFO state and decoded pushbuffer entries.}, label={lst:decoded}]
Doorbell hit, pid 219092
value 0x10011, Kernel Channel 0xFF4A64B8958C3808
=====   GPFIFO SUMMARY   =====
GP_GET      (index)   : 1
GP_PUT      (index)   : 2
GP_base     (VA)      : 0x20021b000
GP_NEWENTRY (VA)      : 0x20021b008
GP_NEWENTRY           : 0x00003e0202600020
===== END GPFIFO SUMMARY =====
Pushbuffer Entries count 15 
PB entry[0] = 0x20048100
    PB HDR INC    count=4 subch=4 addr_dw=0x100 (byte 0x400)
PB entry[1] = 0x00007fa8
    SUBCH4 AMPERE_DMA_COPY_B(0xc7b5) OFFSET_IN_UPPER(0x400) data=0x00007fa8
PB entry[2] = 0x20000000
   SUBCH4 AMPERE_DMA_COPY_B(0xc7b5) OFFSET_IN_LOWER(0x404) data=0x20000000
PB entry[3] = 0x00007fa8
   SUBCH4 AMPERE_DMA_COPY_B(0xc7b5) OFFSET_OUT_UPPER(0x408) data=0x00007fa8
PB entry[4] = 0x0e000000
   SUBCH4 AMPERE_DMA_COPY_B(0xc7b5) OFFSET_OUT_LOWER(0x40c) data=0x0e000000
PB entry[5] = 0x20018106
    PB HDR INC    count=1 subch=4 addr_dw=0x106 (byte 0x418)
PB entry[6] = 0x04000000
    SUBCH4 AMPERE_DMA_COPY_B(0xc7b5) LINE_LENGTH_IN(0x418) data=0x04000000
PB entry[7] = 0x200180c0
    PB HDR INC    count=1 subch=4 addr_dw=0xc0 (byte 0x300)
PB entry[8] = 0x00000182
    SUBCH4 AMPERE_DMA_COPY_B(0xc7b5) LAUNCH_DMA(0x300) data=0x00000182
        DATA_TRANSFER_TYPE=2 (NON_PIPELINED)
        FLUSH_ENABLE=0 (FALSE)
        SRC_MEMORY_LAYOUT=1 (PITCH)
        DST_MEMORY_LAYOUT=1 (PITCH)
        MULTI_LINE_ENABLE=0 (FALSE)
        SRC_TYPE=0 (VIRTUAL)
        DST_TYPE=0 (VIRTUAL)
...
\end{lstlisting}

\subsection{Customizing Command submission}
\label{subsec:customizecmd}
Now that we understand the submission path, we can adjust submitted commands in controlled ways to target specific mechanisms within the hardware and driver and measure the GPU’s response.

In the presence of UVM, GPU virtual addresses are consistent with the process’s user-space virtual addresses (\findingref{finding:GPU_VA}). We therefore track user-space virtual addresses returned by mmap and compare them against the addresses that appear later in our intercepted submission path, including the pushbuffer, GPFIFO, and memory semaphore addresses decoded from command streams. When a match is observed, we can attribute that mmap allocation to the corresponding object (pushbuffer, GPFIFO, and semaphore buffer).

After identifying these allocations, we inject customized commands by writing directly to the identified pushbuffer and GPFIFO, then ring the doorbell by writing the 32-bit channel ID. For time measurement, we read and calculate timestamps written by GPU from the semaphore buffer.

\section{Case Studies: Unveiling Driver Logic and Optimization}
\label{sec:casestudy}
Using the methodology introduced in \Cref{sec:method}, we present two case studies that show how our modified driver can separate hardware and software side effects.

First, we study cudaMemcpy from a hardware-centric perspective. By issuing a customized command sequence, we directly measure GPU DMA engine behavior excluding driver-side overheads that can obscure short transfers. Second, we study CUDA Graphs from a software-centric perspective. We compare graph execution under CUDA 11.8 and CUDA 13.0, and use the reconstructed command streams to pinpoint driver-side changes that align with the observed reduction in host launch cost.

\subsection{Evaluation platform}
\label{subsec:evaluation_platform}
The hardware of our evaluation platform is shown in \Cref{tab:hw_spec_sub}, and software stacks used in our experiments are summarized in \Cref{tab:software_stacks_sub}. We evaluate two major stacks, based on CUDA 11.8 and CUDA 13.0. For all performance measurements, we use the unmodified OpenGPU kernel module, denoted as 11.8-perf and 13.0-perf. For command logging, we use a modified OpenGPU driver, denoted as 11.8-log and 13.0-log.

To run the modified driver, we export \texttt{sched\_task\_fork} from the default Linux 5.14 kernel to offload sleepable operations out of the watchpoint handler.


While we evaluate a single GPU platform, the GPU driver submission workflow (pushbuffer, GPFIFO filling and doorbell triggering) remains the same across generations. Therefore, one platform is sufficient to demonstrate the methodology, and adapting it to other models only requires altering to the header of the given model.

\begin{table}[h]
\centering
\caption{Experimental platform}
\label{tab:platform_and_stack}

\begin{subtable}[h]{\linewidth}
\centering
\caption{Hardware configuration.}
\label{tab:hw_spec_sub}
\begin{tabular}{@{}l l@{}}
\toprule
CPU & Intel Xeon Gold 6338 @ 2.00\,GHz \\
GPU & NVIDIA A40 (Ampere) \\
Interconnect & PCIe Gen4 $\times$16 (16\,GT/s capable) \\
\bottomrule
\end{tabular}
\end{subtable}
\hfill
\vspace{4pt}
\begin{subtable}[h]{\linewidth}
\centering
\caption{Software stacks.}
\small
\label{tab:software_stacks_sub}
\begin{tabular}{@{}lcccc@{}}
\toprule
\textbf{Component} & \textbf{11.8-perf} & \textbf{11.8-log} & \textbf{13.0-perf} & \textbf{13.0-log}\\ 
\midrule
CUDA Toolkit & \multicolumn{2}{c}{11.8} & \multicolumn{2}{c}{13.0} \\
\makecell[l]{Userspace driver} & \multicolumn{2}{c}{520.61.07}  & \multicolumn{2}{c}{580.105.08} \\
\makecell[l]{Opengpu kernel\\module} & \makecell[c]{520.61.07 \\ (original)} & \makecell[c]{520.61.07 \\ (modified)} & \makecell[c]{580.105.08 \\(original)}  & \makecell[c]{580.105.08 \\ (modified)} \\
Operating System & \multicolumn{4}{c}{Rocky Linux 9.5} \\
Linux Kernel & \multicolumn{4}{c}{5.14 (Patched)} \\
\bottomrule
\end{tabular}
\end{subtable}

\end{table}

\begin{findingbanner}[title=Finding 2: GPFIFO and pushbuffer locality]
\label{finding:locality}
In our baremetal environment, we find that the
GPFIFO ring buffer resides in GPU video memory, while the pushbuffer is
allocated in host RAM. This creates an asymmetric submission path: the CPU
writes pushbuffer commands locally in system memory and then updates the GPFIFO
entries remotely on the GPU, whereas the GPU performs the reverse---it reads
GPFIFO entries locally from its own RAM and then fetches pushbuffer commands
remotely from host memory.
\end{findingbanner}

\begin{figure}[h]
  \centering

  \begin{subfigure}[h]{\columnwidth}
    \centering
    \includegraphics[width=0.8\linewidth]{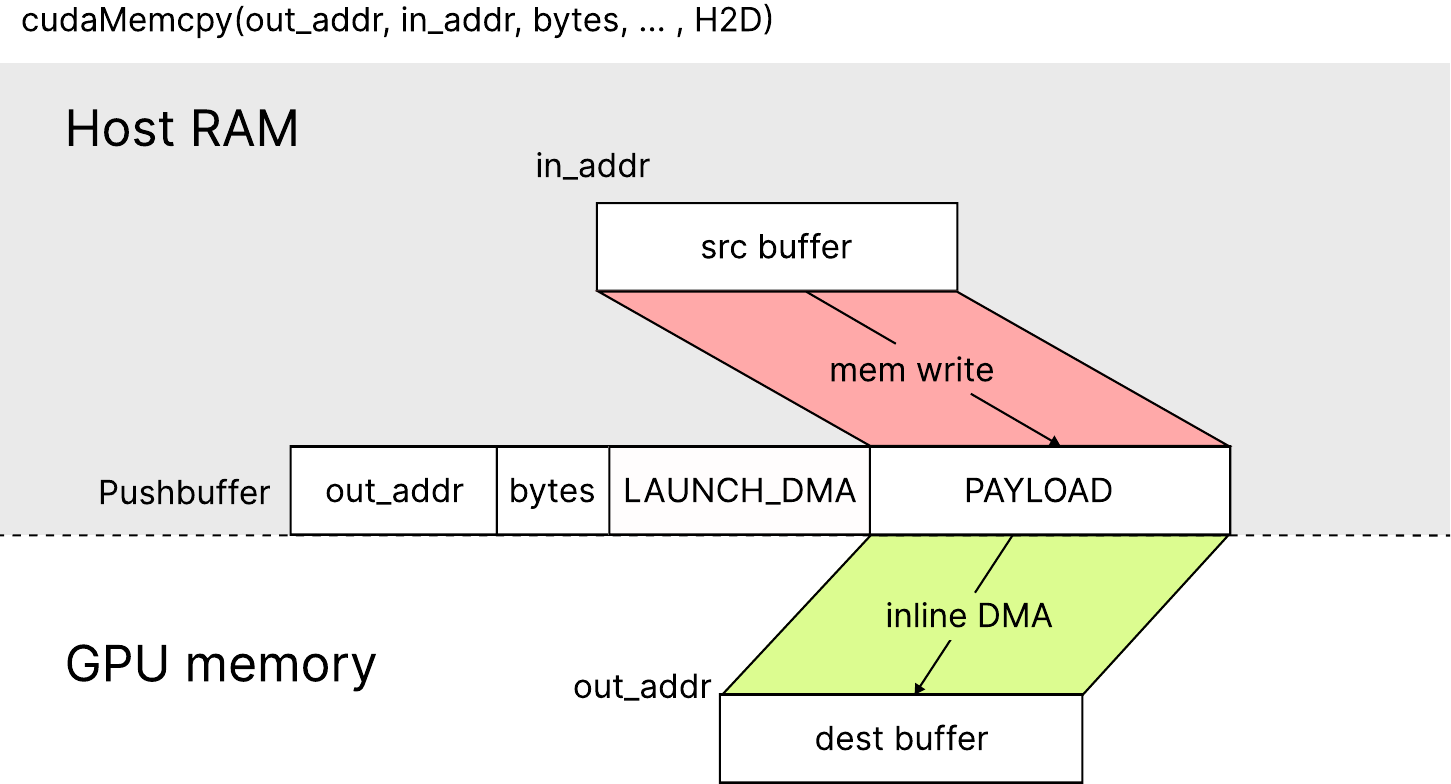}
    \caption{Inline DMA.}
    \label{fig:inlinedma}
  \end{subfigure}

  \vspace{0.6em}

  \begin{subfigure}[h]{\columnwidth}
    \centering
    \includegraphics[width=0.8\linewidth]{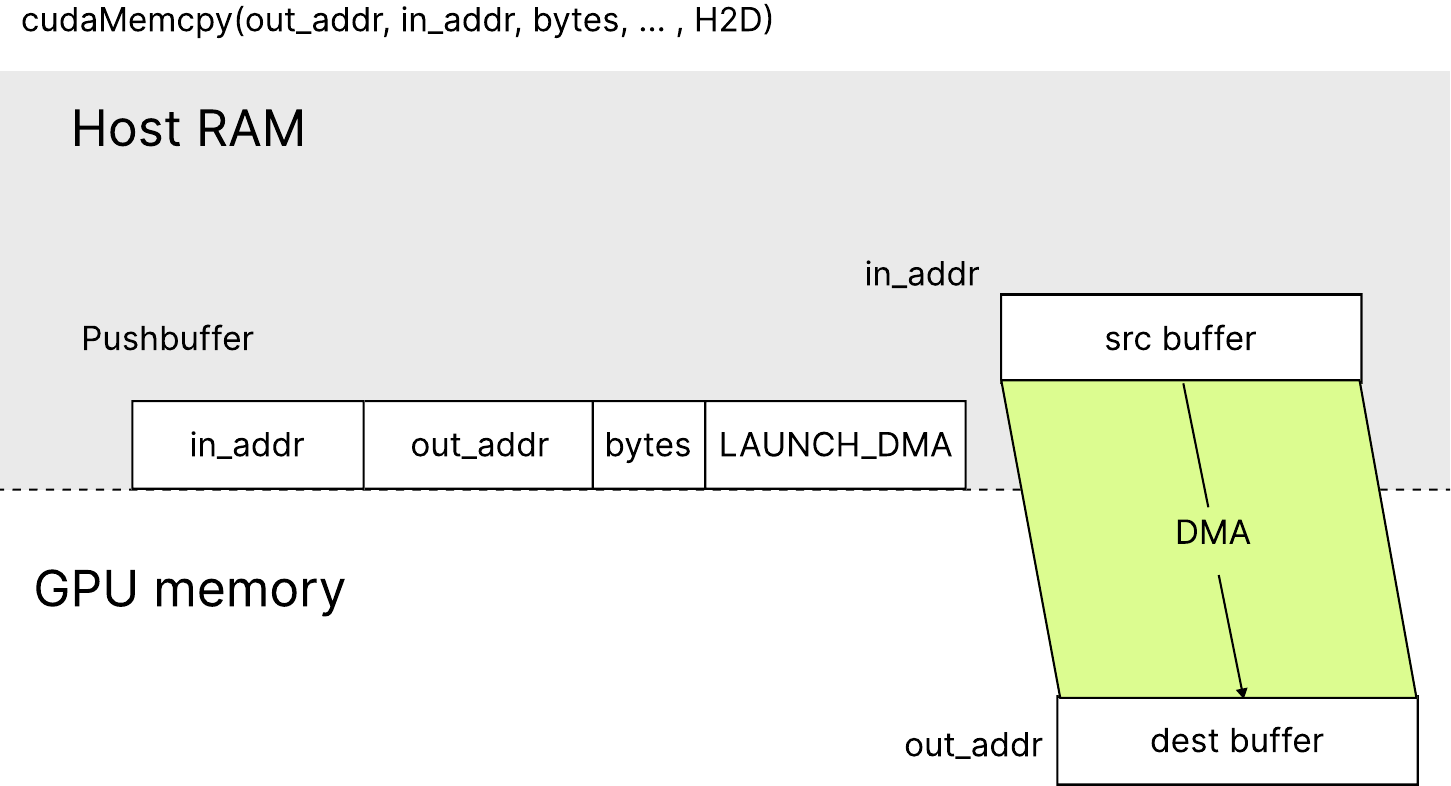}
    \caption{Direct DMA.}
    \label{fig:directdma}
  \end{subfigure}

  \caption{DMA submission paths of CUDA data movement}
  \label{fig:dma_paths}
\end{figure}

\subsection{Analyzing CUDA Data Movement Mechanism}
We begin by analyzing \texttt{cudaMemcpy}, a CUDA API that explicitly transfers data between different domains. A \texttt{cudaMemcpy} operation is defined by the source address, destination address, transfer direction, and transfer size. Because it is the explicit data movement interface in CUDA, it is commonly used to measure transfer latency and bandwidth.

In the Host-to-Device direction of \texttt{cudaMemcpy}  on the 13.0-log stack (CUDA 13.0), we observe two distinct command-flow patterns. In one case, the transfer is issued using an inline DMA (Direct Memory Access) path. In the other, it follows a direct DMA path. We illustrate these two patterns in \Cref{fig:inlinedma} and \Cref{fig:directdma}.



When the transfer size is small (< 24\,KiB on our environment), the driver uses the inline DMA method. In this mode, the pushbuffer specifies only the destination address (\texttt{out\_addr}) and the transfer size, while the source data is embedded directly as part of the remaining pushbuffer payload. On the device side, the \textbf{compute engine} fetches this staged payload and writes it to the destination.

For larger transfers ($\geq$ 24\,KiB), the driver switches to the direct DMA method. Here, the pushbuffer command explicitly specifies both the source address and destination address, and the transfer is executed by a dedicated \textbf{copy engine} rather than the compute engine. 

\subsubsection*{Extracting the hardware performance of DMA engines}
While \\ \texttt{cudaMemcpy} is widely used to evaluate interconnect capability, several factors prevent it from directly exposing the underlying hardware behavior.

First, as shown above for host-to-device transfers, two distinct DMA mechanisms exist: inline DMA through the compute engine and direct DMA through the copy engine. These mechanisms follow different submission paths and exhibit different performance characteristics. Since the CUDA runtime does not expose control over which mechanism is used, the CUDA API alone does not permit a direct comparison between them.

Second, timing accuracy presents another obstacle. A \texttt{cudaMemcpy} call issues one transfer at a time, so the measured latency includes more than the engine transfer time alone. Since runtime-level timing does not isolate the engine execution boundary itself, the reported duration also includes part of the submission path before the transfer begins.


Using the command customization techniques in \Cref{subsec:customizecmd}, we could explicitly control transfer size and DMA mode, providing a complete map of behavior across size–mode combinations. In our benchmarking, we implement this by coalescing all transfer commands and their progress trackers covering both the warm-up phase and the measured phase into a single pushbuffer segment. The command segment is organized as (\allowbreak\texttt{transfer\allowbreak\_cmd}\allowbreak\,\allowbreak$\times$\allowbreak\,\allowbreak\texttt{warmup\allowbreak\_iters}\allowbreak),\allowbreak\ \texttt{warmup\allowbreak\_tracker}\allowbreak,\allowbreak\
(\allowbreak\texttt{transfer\allowbreak\_cmd}\allowbreak\,\allowbreak$\times$\allowbreak\,\allowbreak\texttt{test\allowbreak\allowbreak\_iters}\allowbreak),\allowbreak\ \texttt{test\allowbreak\_tracker}\allowbreak
, and we submit this entire command stream once. After this submission, no further driver intervention occurs while the GPU runs through the sequence uninterrupted. We then poll on the two progress trackers until the expected payload values are observed, read out the corresponding \mbox{timestamps}, and subtract them. As a result, the repeated transfers execute without further host-side or driver-side intervention, and the elapsed time is determined entirely from device-side start and completion timestamps, more directly reflecting the underlying hardware performance.



\subsubsection*{Performance Analysis of two DMA modes}
\label{subsubsec:dmabwlat}

\begin{figure}[htb]
\centering
\captionsetup{justification=centering}

\begin{subfigure}{0.48\columnwidth}
  \centering
  \includegraphics[width=\linewidth]{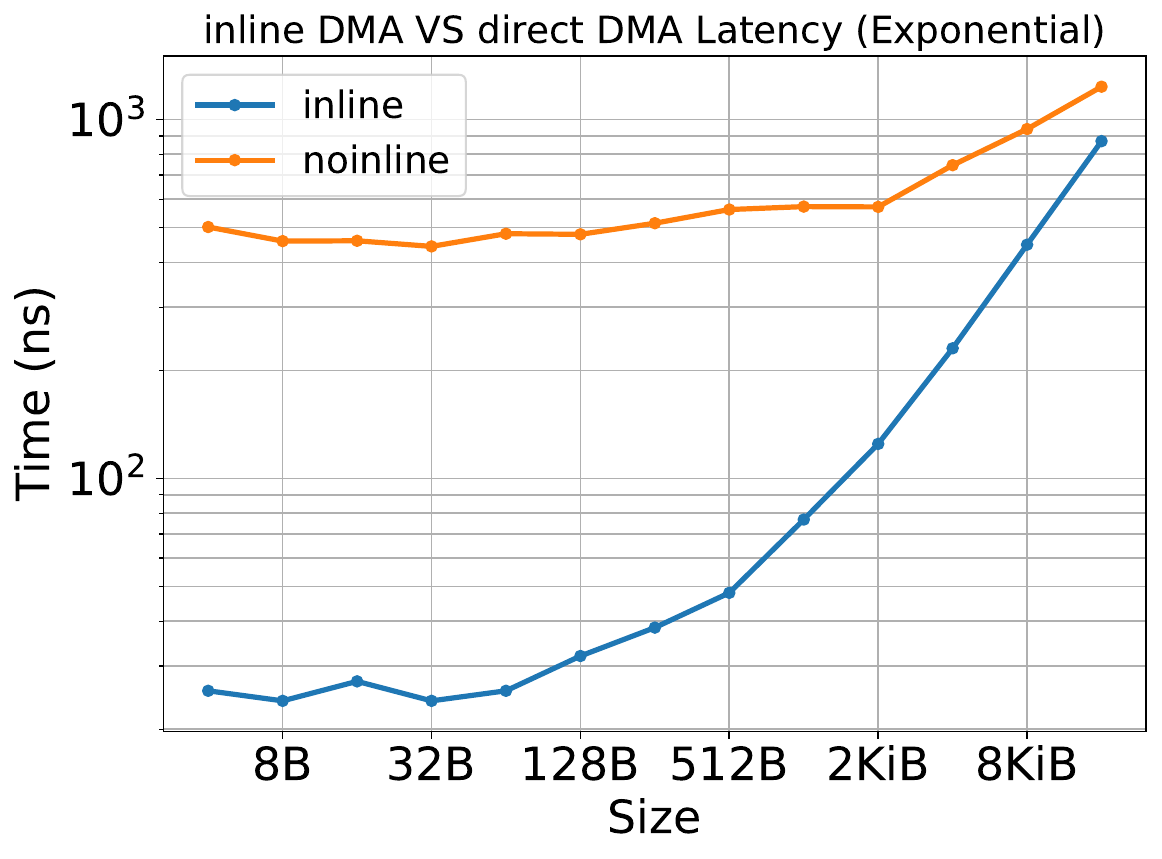}
  \caption{Latency(ns), size grow exponentially from 4B to 16KiB}
  \label{fig:time_exp}
\end{subfigure}\hfill
\begin{subfigure}{0.48\columnwidth}
  \centering
  \includegraphics[width=\linewidth]{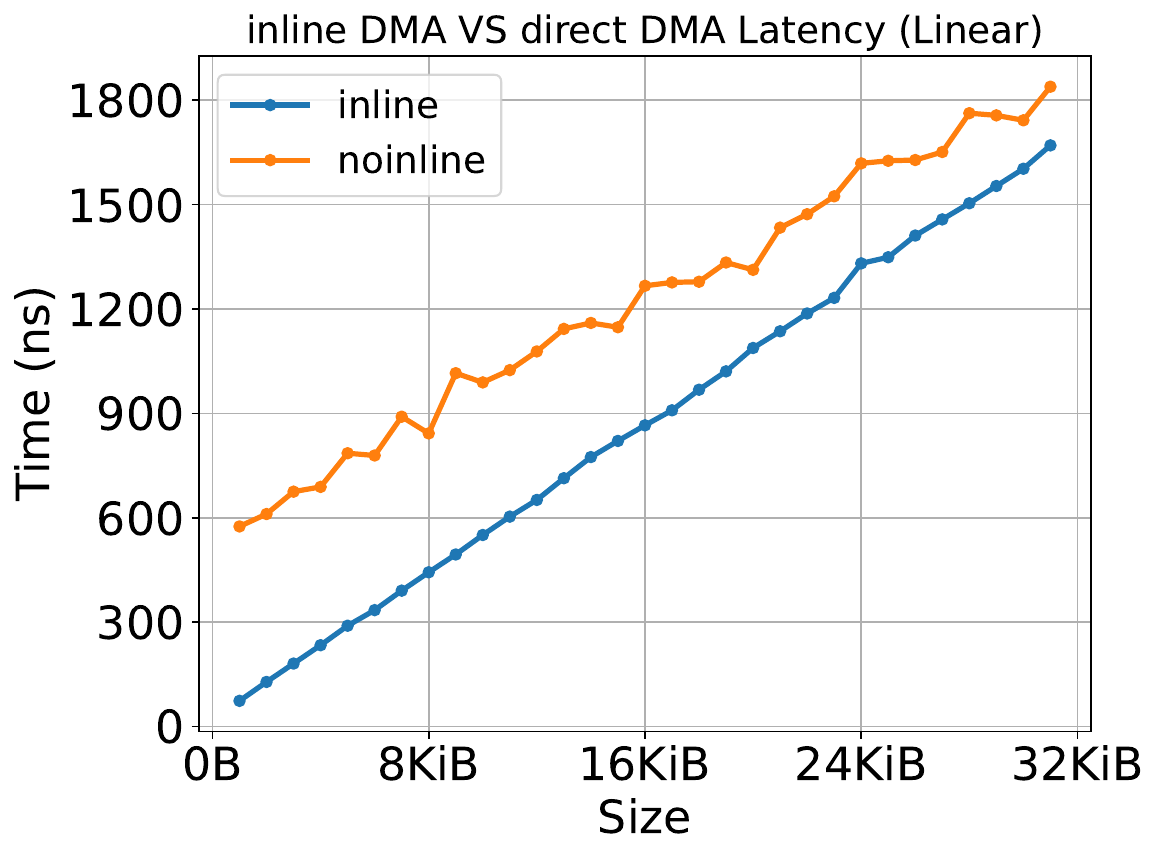}
  \caption{Latency(ns), size grow linearly from 1KiB to 31KiB}
  \label{fig:time_linear}
\end{subfigure}

\vspace{0.4em}

\begin{subfigure}{0.50\columnwidth}
  \centering
  \includegraphics[width=\linewidth]{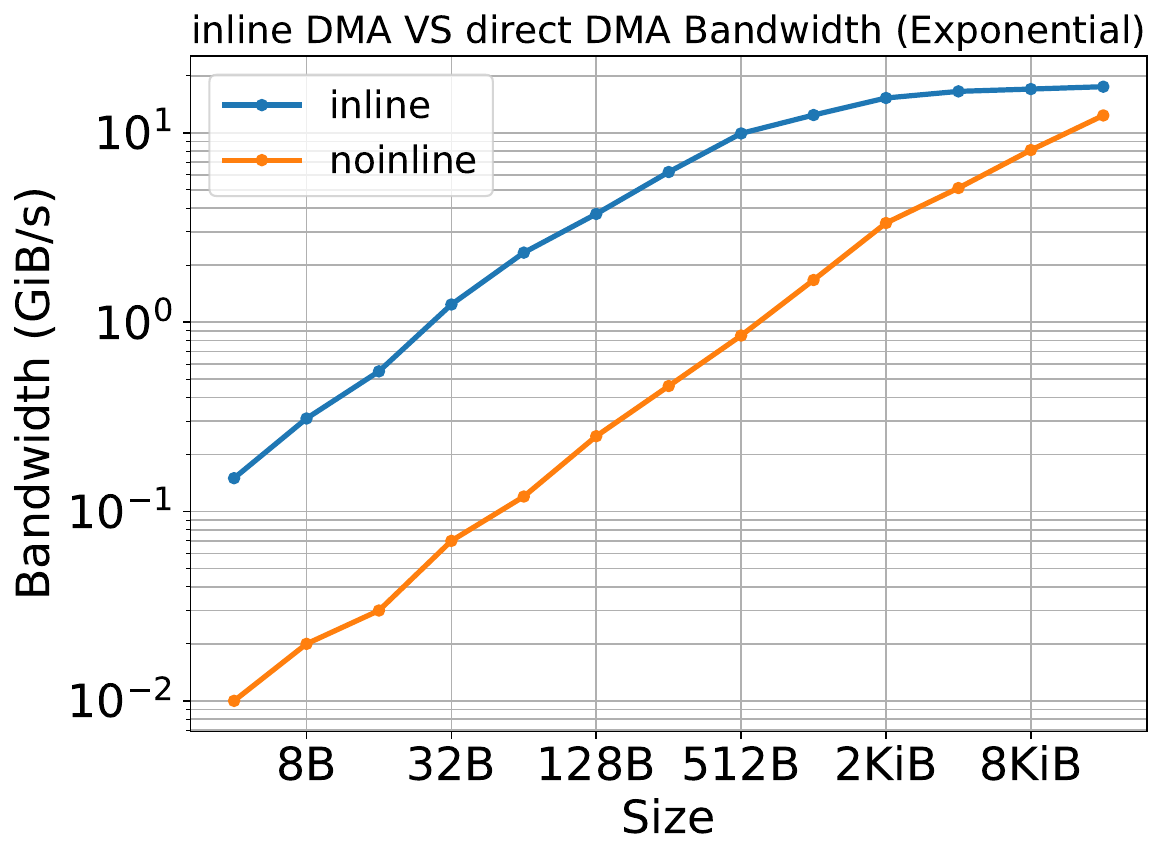}
  \caption{Bandwidth(GiB/s), size grow exponentially from 4B to 16KiB}
  \label{fig:Bw_exp}
\end{subfigure}\hfill
\begin{subfigure}{0.50\columnwidth}
  \centering
  \includegraphics[width=\linewidth]{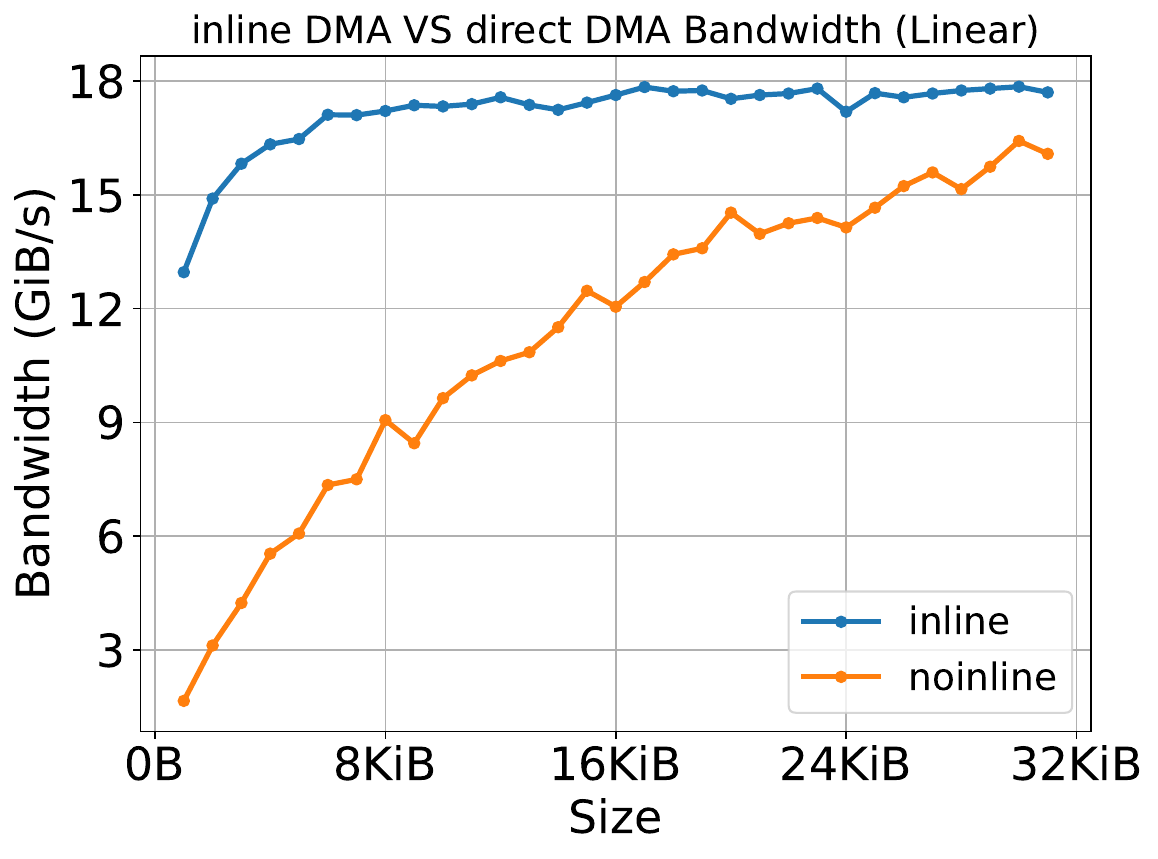}
  \caption{Bandwidth(GiB/s), size grow linearly from 1KiB to 31KiB}
  \label{fig:Bw_linear}
\end{subfigure}

\caption{DMA transfer performance across two copy engines. Top: Latency, Bottom: Bandwidth}
\label{fig:dmaperf}
\end{figure}


We plot the performance of the two DMA submission modes in \Cref{fig:dmaperf}. The top row reports latency (\Cref{fig:time_exp,fig:time_linear}), and the bottom row reports bandwidth (\Cref{fig:Bw_exp,fig:Bw_linear}). We evaluate two transfer-size sweeps: an exponential sweep from \SI{4}{\byte}, doubling up to \SI{16}{\kibi\byte}, and a linear sweep from \SI{1}{\kibi\byte} to \SI{31}{\kibi\byte}. We cap the linear sweep at \SI{31}{\kibi\byte}, because transfers larger than this were not accepted by the compute-engine in our experiments.

The results show that inline DMA on the compute engine achieves a much lower startup latency of about \SI{24}{\nano\second}, compared with about \SI{500}{\nano\second} for copy engine. However, the compute-engine path saturates quickly, reaching about 17.5~GiB/s at a transfer size of \SI{8}{\kibi\byte}. The copy-engine in contrast scales to larger transfer sizes and reaches its saturation bandwidth of 22~GiB/s at around \SI{1}{\mebi\byte} in our experiments, which is beyond the range shown in the figure.

The latency extracted from raw hardware performance shows a clear disparity from the Nsight "CUDA HW" duration (\Cref{tab:nsightvours}). The left half of the table corresponds to the compute-engine , while the right half corresponds to the copy-engine(the protocol switch is at \SI{24}{\kibi\byte} in our environment). The percentage column is computed as $(T_{\mathrm{Nsight}}-T_{\mathrm{raw}})/T_{\mathrm{Nsight}}$, representing the fraction of profiler-reported latency not accounted for by raw hardware execution.

This fraction shows a clear declining trend as transfer size grows, since hardware transmission time increasingly dominates. As discussed earlier in this section, the compute-engine and copy-engine paths involve different mechanisms, so the non-hardware portion has different meanings in the two cases. For the compute-engine path, NSight could include 4 things in its measurements, driver-side staging of user data into the command buffer, PBDMA fetching of that command buffer, compute engine loading the inlined data and Nsight overheads. However, the documentation does not clearly define what is included in this interval. Although Nsight labels it as ``CUDA HW'', it does not expose how the reported time is partitioned internally. Thus, the source of the extra time seen by Nsight could combine runtime-level submission/measurement overhead with CPU-side staging work for inlined data. By separating the raw engine execution time, our method clarifies the attribution and enables further decomposition of the remaining stages.

For the copy-engine path, the transfer itself is carried out by the copy engine between host and GPU memory. Therefore, the gap between Nsight-reported latency and raw DMA latency mainly reflects runtime and submission overhead outside engine execution. Our method removes this runtime-level offset and more accurately reflects the underlying engine behavior.

\begin{table}[h]
\centering
\setlength{\tabcolsep}{1pt}
\renewcommand{\arraystretch}{1.05}
\caption{Nsight-measured latency VS raw DMA latency.}
\label{tab:nsightvours}
\begin{tabular}{c c c c | c c c c}
\hline
\multicolumn{4}{c|}{Compute engine} & \multicolumn{4}{c}{Copy engine} \\
\multicolumn{4}{c|}{(Inline DMA)} & \multicolumn{4}{c}{(Non-inline/direct DMA)} \\
\cline{1-4}\cline{5-8}
\hline
Size & Nsight(ns) & raw(ns) & \%
     & Size & Nsight($\mu$s) & raw($\mu$s) & \% \\
\hline
8    & 468.25  & 24.00  & 94.87\%
     & 32 Ki  & 3.78    & 1.90    & 49.89\% \\
32   & 474.50  & 24.00  & 94.94\%
     & 128 Ki & 6.97    & 5.95    & 14.65\% \\
128  & 495.50  & 32.00  & 93.54\%
     & 512 Ki & 22.80   & 22.06   & 3.25\% \\
512  & 564.50  & 48.00  & 91.50\%
     & 2 Mi   & 87.89   & 87.11   & 0.89\% \\
2 Ki & 1763.50 & 124.80 & 92.92\%
     & 8 Mi   & 348.60  & 346.90  & 0.49\% \\
8 Ki & 1924.75 & 448.00 & 76.72\%
     & 32 Mi  & 1389.98 & 1384.96 & 0.36\% \\
\hline
\end{tabular}
\end{table}



\subsection{Evolution of CUDA Graph from 11.8 to 13.0}

\begin{figure}[t]
\centering

\begin{subfigure}[t]{0.48\linewidth}
  \captionsetup{justification=centering}
  \centering
  \includegraphics[width=\linewidth]{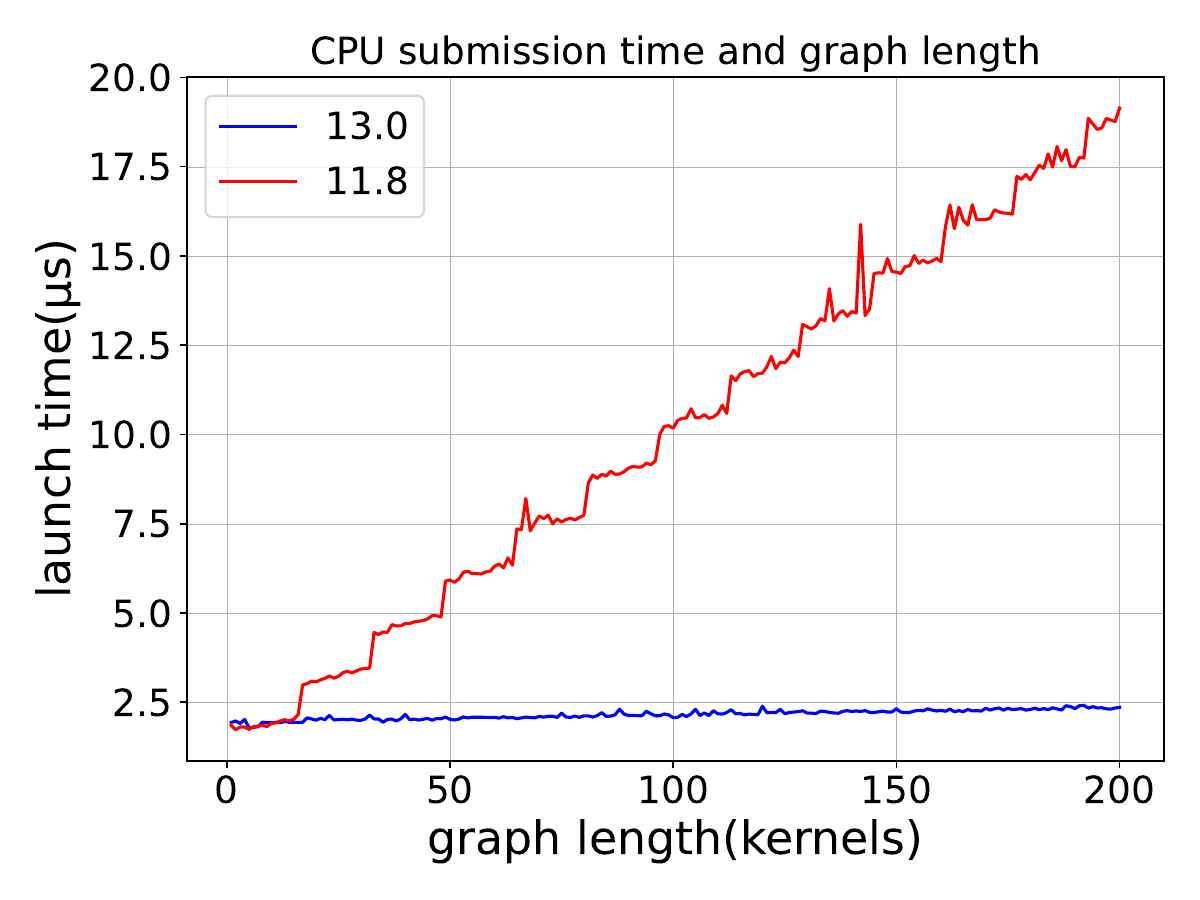}
  \caption{CPU submission (launch) time vs.\ graph length (0--200).}
  \label{fig:chainlen_vs_time_0_200}
\end{subfigure}\hfill
\begin{subfigure}[t]{0.48\linewidth}
  \captionsetup{justification=centering}
  \centering
  \includegraphics[width=\linewidth]{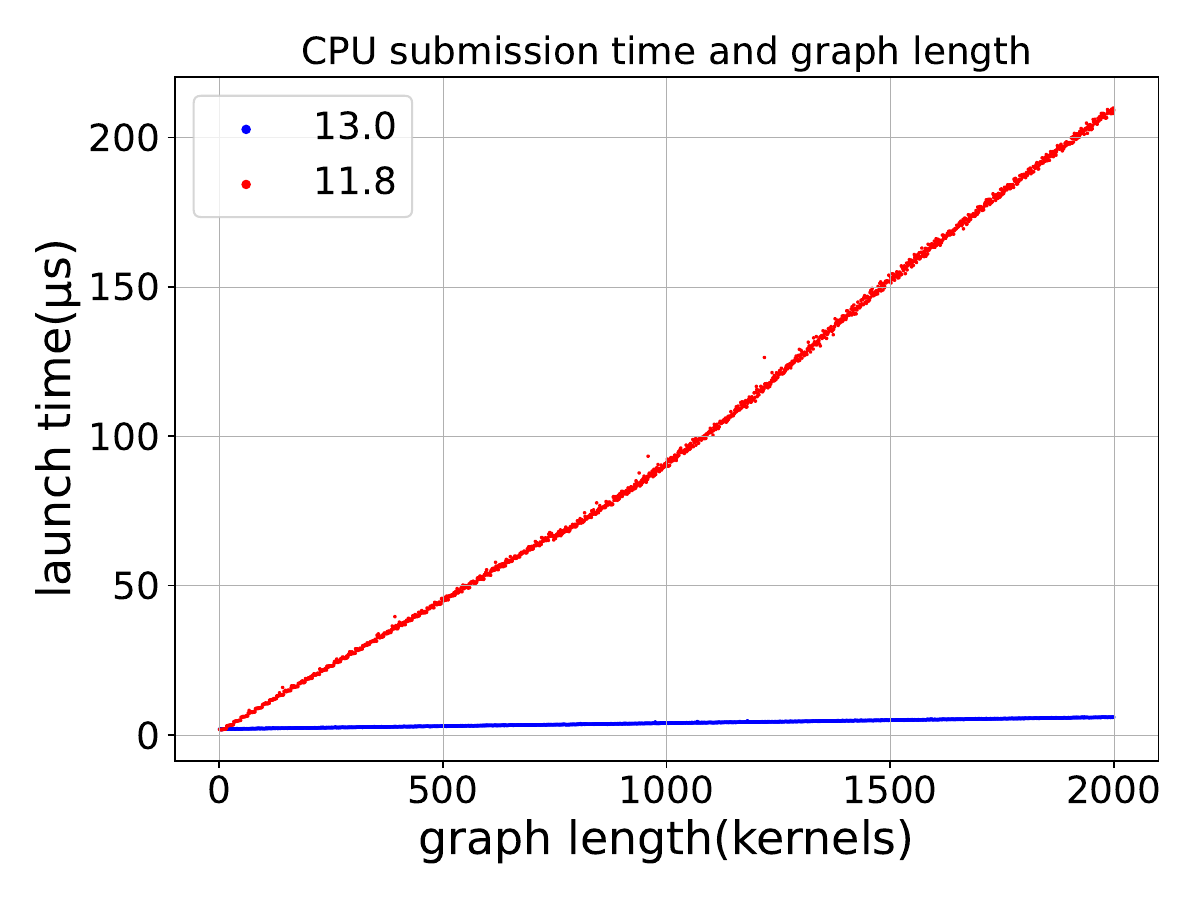}
  \caption{CPU submission (launch) time vs.\ graph length (0--2000).}
  \label{fig:chainlen_vs_time_0_2000}
\end{subfigure}

\vspace{0.6em}

\begin{subfigure}[t]{0.48\linewidth}
  \captionsetup{justification=centering}
  \centering
  \includegraphics[width=\linewidth]{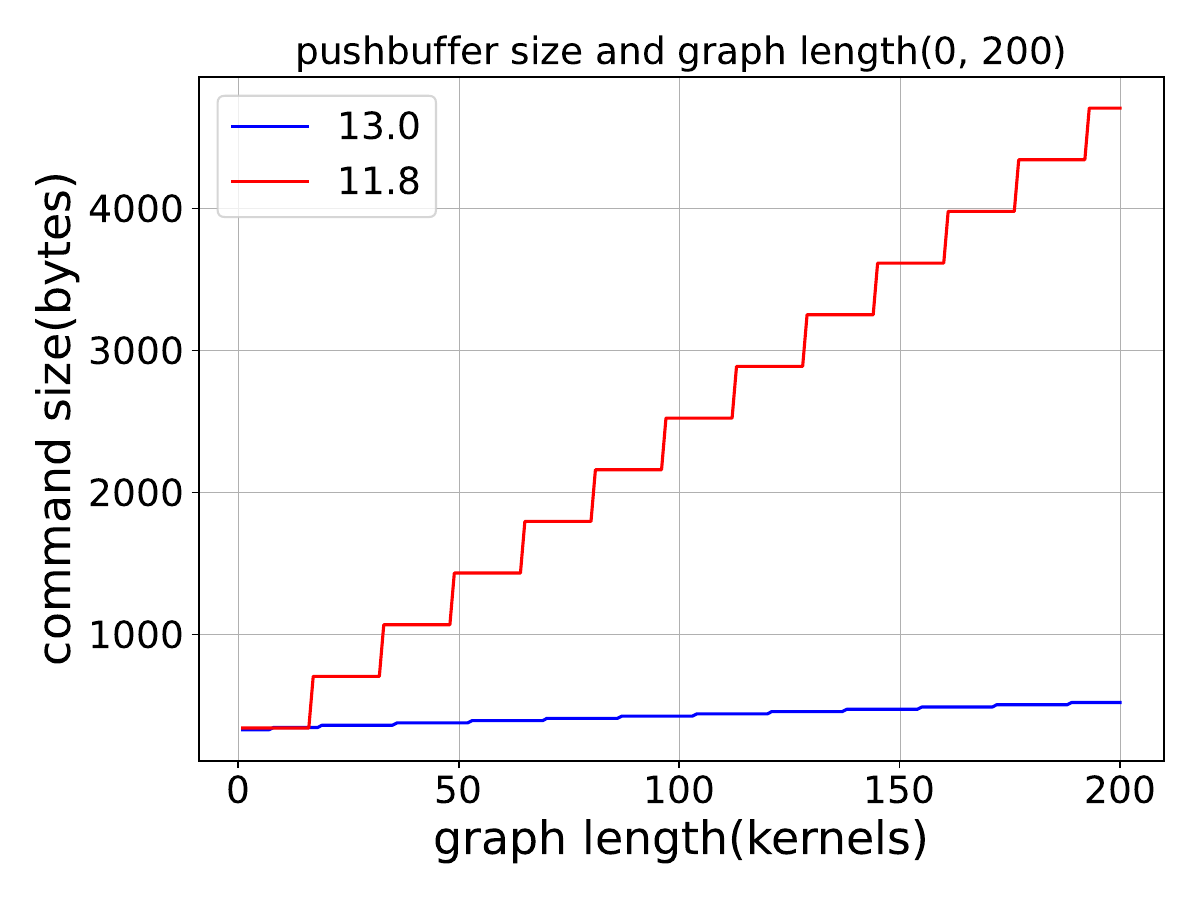}
  \caption{Command size vs.\ graph length (0--200).}
  \label{fig:chainlen_vs_size_0_200}
\end{subfigure}\hfill
\begin{subfigure}[t]{0.48\linewidth}
  \captionsetup{justification=centering}
  \centering
  \includegraphics[width=\linewidth]{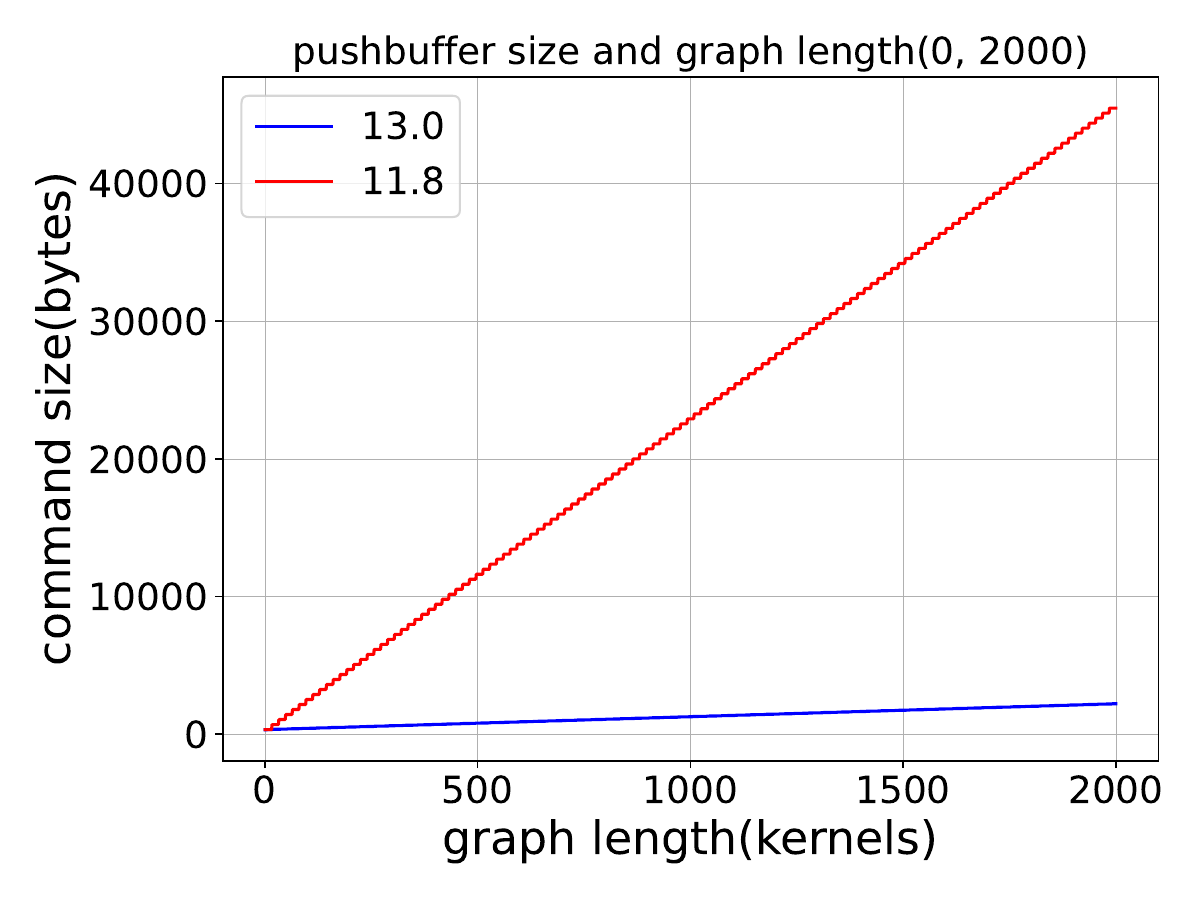}
  \caption{Command size vs.\ graph length (0--2000).}
  \label{fig:chainlen_vs_size_0_2000}
\end{subfigure}

\vspace{0.6em}

\begin{subfigure}[t]{0.48\linewidth}
  \captionsetup{justification=centering}
  \centering
  \includegraphics[width=\linewidth]{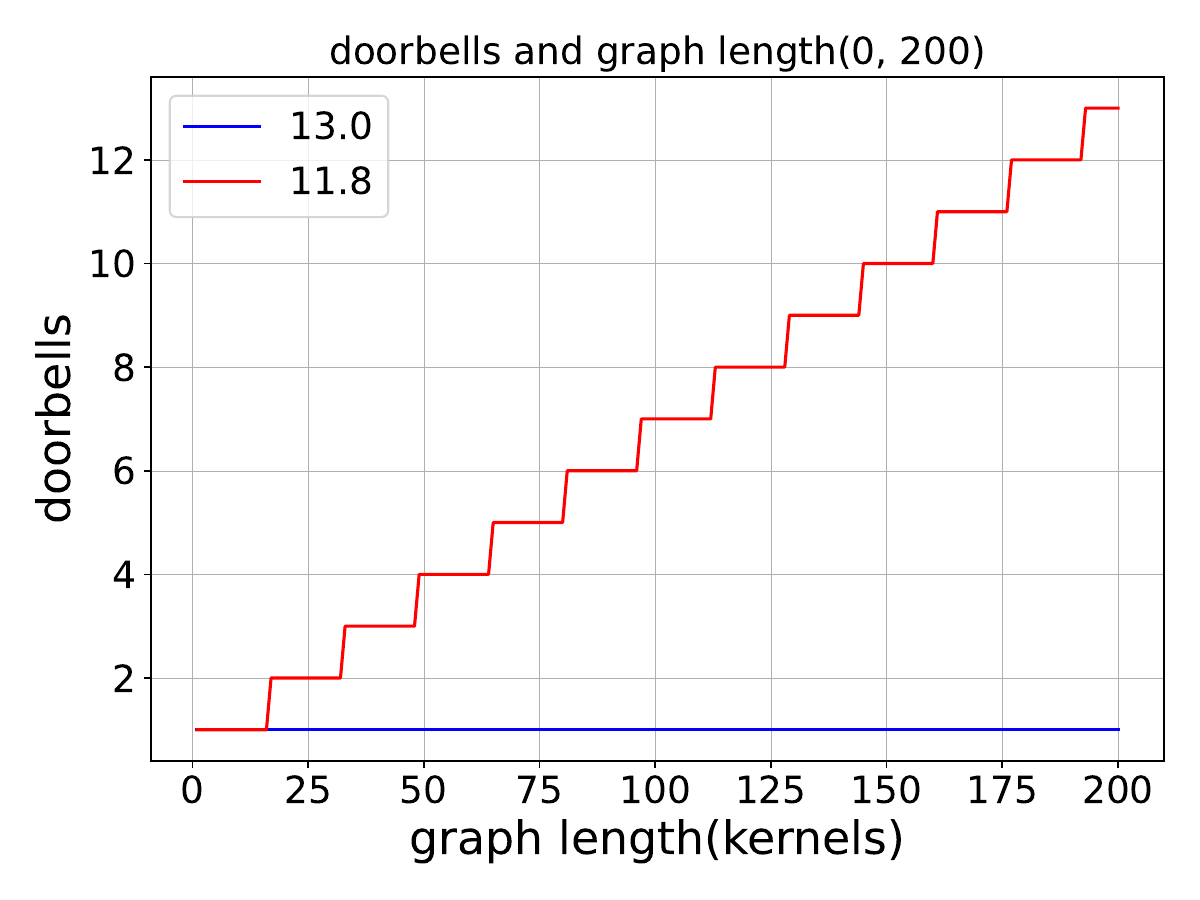}
  \caption{Doorbell writes vs.\ graph length (0--200).}
  \label{fig:chainlen_vs_doorbells_0_200}
\end{subfigure}\hfill
\begin{subfigure}[t]{0.48\linewidth}
  \captionsetup{justification=centering}
  \centering
  \includegraphics[width=\linewidth]{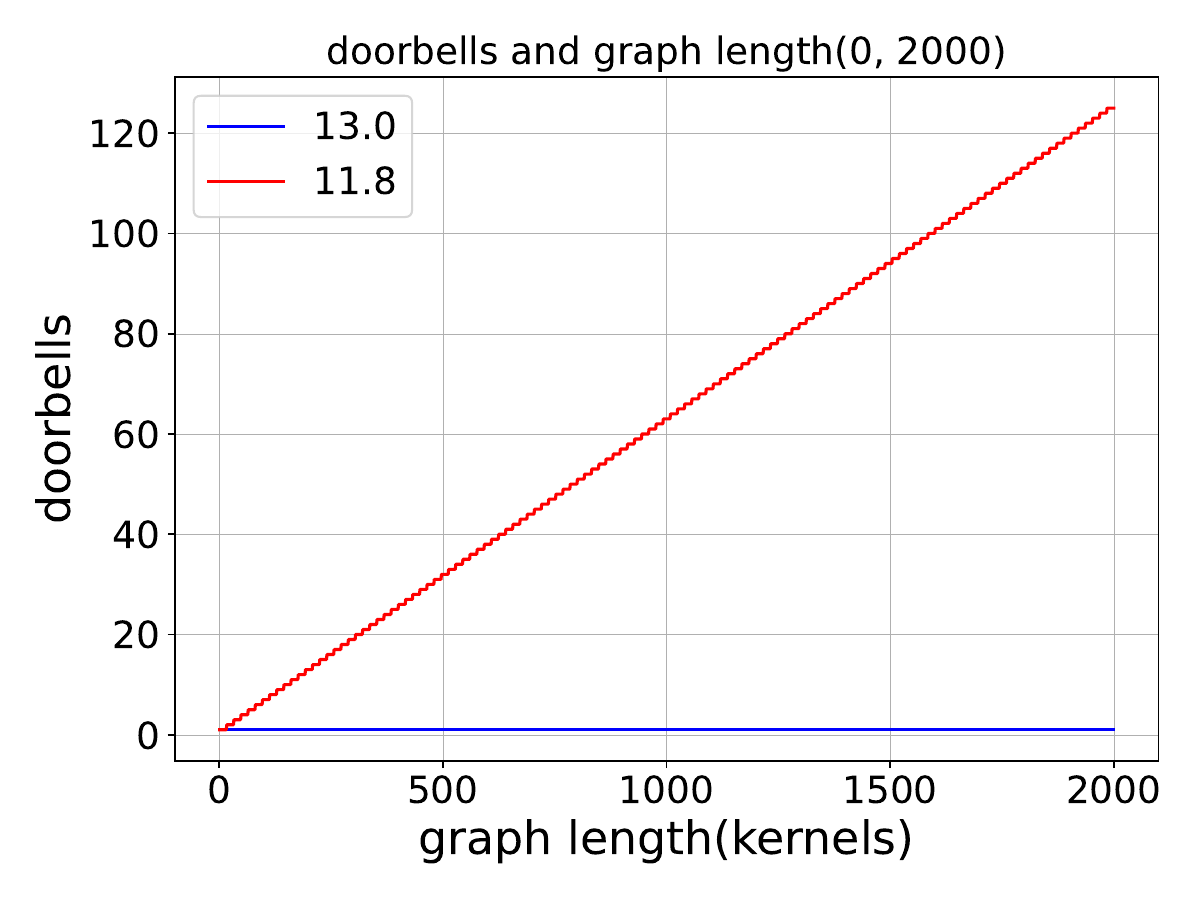}
  \caption{Doorbell writes vs.\ graph length (0--2000 kernels).}
  \label{fig:chainlen_vs_doorbells_0_2000}
\end{subfigure}

\captionsetup{justification=raggedright,singlelinecheck=false}
\caption{Scaling of CPU submission cost and command stream activity with CUDA Graph chain length.
Top to bottom: CPU submission time, command size, and doorbell writes. Left: short chains (0--200), right: full range to 2000.}
\label{fig:submission_vs_all}
\end{figure}


\begin{figure}[h]
  \centering
  \includegraphics[width=0.85\linewidth]{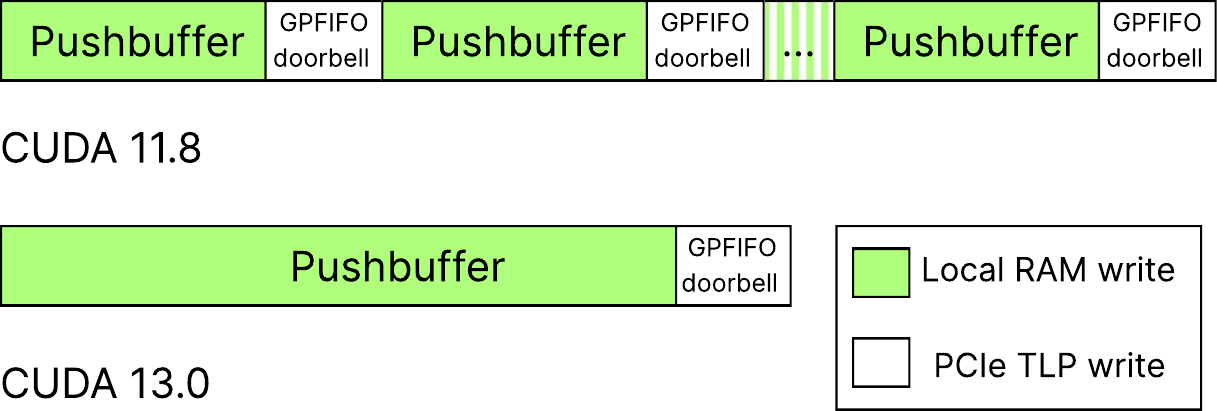}
  \caption{Submission patterns of CUDA 11.8 (top) and CUDA 13.0 (bottom).}
  \label{fig:pattern}
\end{figure}

Next, we study CUDA Graph execution in detail. 
CUDA Graph allows a CUDA program to represent a sequence of operations, such as kernel launches and memory copies, together with their dependencies. Compared with issuing these operations individually, launching them as a graph can significantly reduce CPU overhead by avoiding repeated API invocation, redundant submission work, and repeated command emission to the GPU. This advantage is especially important when the same sequence is executed many times, such as in AI training loops or time-stepped physics simulations.

In this workflow, \texttt{cudaGraphUpload} uploads reusable execution metadata for the graph, and \texttt{cudaGraphLaunch} subsequently triggers execution using that uploaded state. This separation makes CUDA Graph a useful mechanism for reducing the redundant CPU-side cost of repeatedly launching the same sequence of operations.

Recent advances from CUDA 11.8 to 13.0 for Ampere+ GPUs further reduce graph launch overhead~\cite{cudagraph}. Rather than increasing with the number of kernel nodes, the launch cost in CUDA 13.0 remains nearly constant, compared to CUDA 11.8 where launch time grows linearly with kernel count inside the graph.

For our investigation, we first reproduce the performance figures for CUDA 11.8 and CUDA 13.0 using their corresponding driver stacks, \textbf{11.8-perf} and \textbf{13.0-perf}, as described in \Cref{subsec:evaluation_platform}. We then substitute the kernel driver with our modified versions, \textbf{11.8-log} and \textbf{13.0-log}, to trace the command traffic generated under varying graph sizes during benchmarking.

The graph structure we use is a sequence of kernel launches issued to the same stream. Since a CUDA stream enforces in-order execution, each kernel is dependent on the completion of the previous one, yielding a simple \emph{chain} structured graph. We therefore define the graph size (length) as the number of kernels in the chain. We benchmarked for the graph size ranging from 1 to 2000.

Each graph node launches an identical short compute kernel operating on an $N$-element array, which applies a fixed scalar multiplication to each element.


While our tool can recover field names in the GPU command stream, many commands emitted during \texttt{cudaGraphLaunch} use NVIDIA-internal terms whose semantics are not publicly documented. Rather than speculate on individual closed-source fields, we analyze graph execution through submission-level indicators: doorbell activity and reconstructed command volume. These metrics provide a macroscopic but mechanism-relevant view of driver behavior, reflecting both CPU-side command work and interaction frequency with the GPU submission path.


\subsubsection{Command size and launch-time scaling} 
\begin{figure}[htb]
\centering

\begin{subfigure}{0.48\columnwidth}
  \centering
  \includegraphics[width=\linewidth]{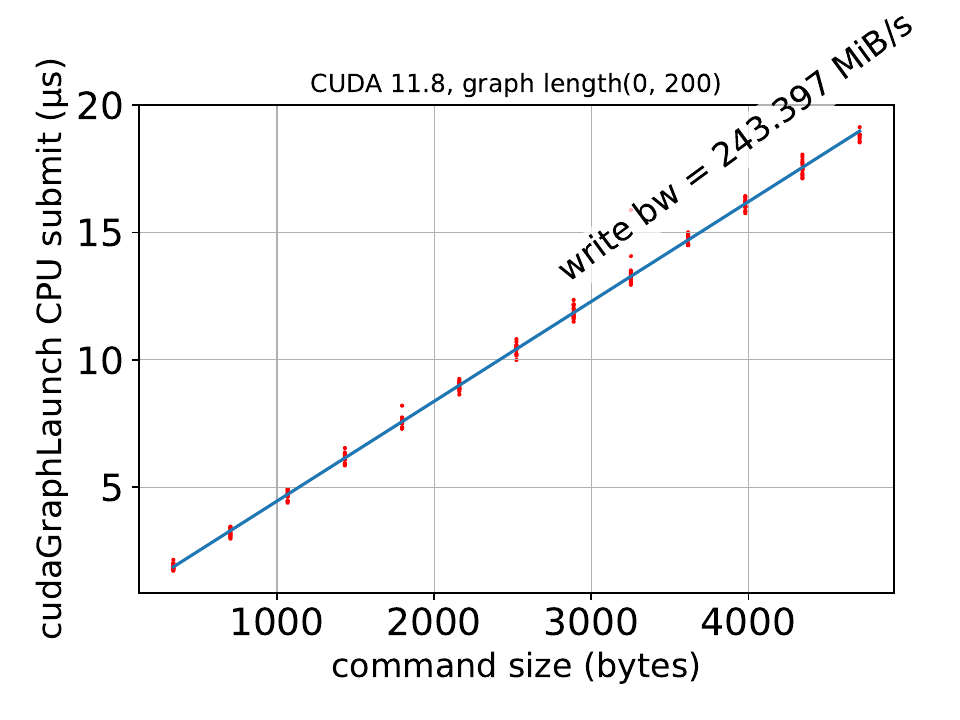}
  \caption{CUDA 11.8, graph length 0–200}
  \label{fig:11_8_pb_vs_time_0_200}
\end{subfigure}\hfill
\begin{subfigure}{0.48\columnwidth}
  \centering
  \includegraphics[width=\linewidth]{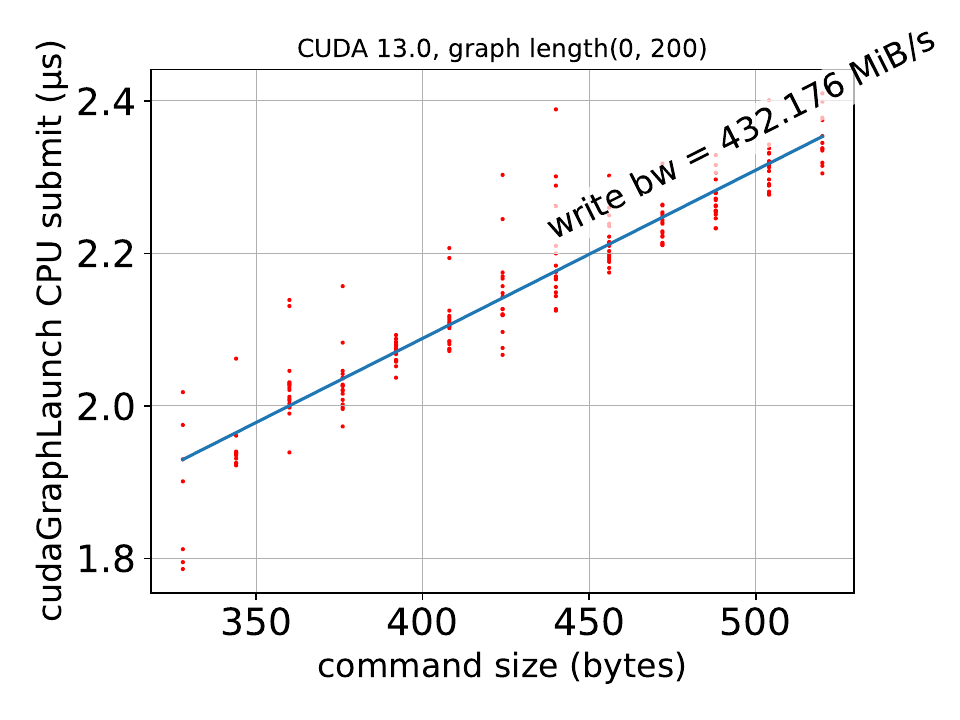}
  \caption{CUDA 13.0, graph length 0–200}
  \label{fig:13_0_pb_vs_time_0_200}
\end{subfigure}

\vspace{0.4em}

\begin{subfigure}{0.50\columnwidth}
  \centering
  \includegraphics[width=\linewidth]{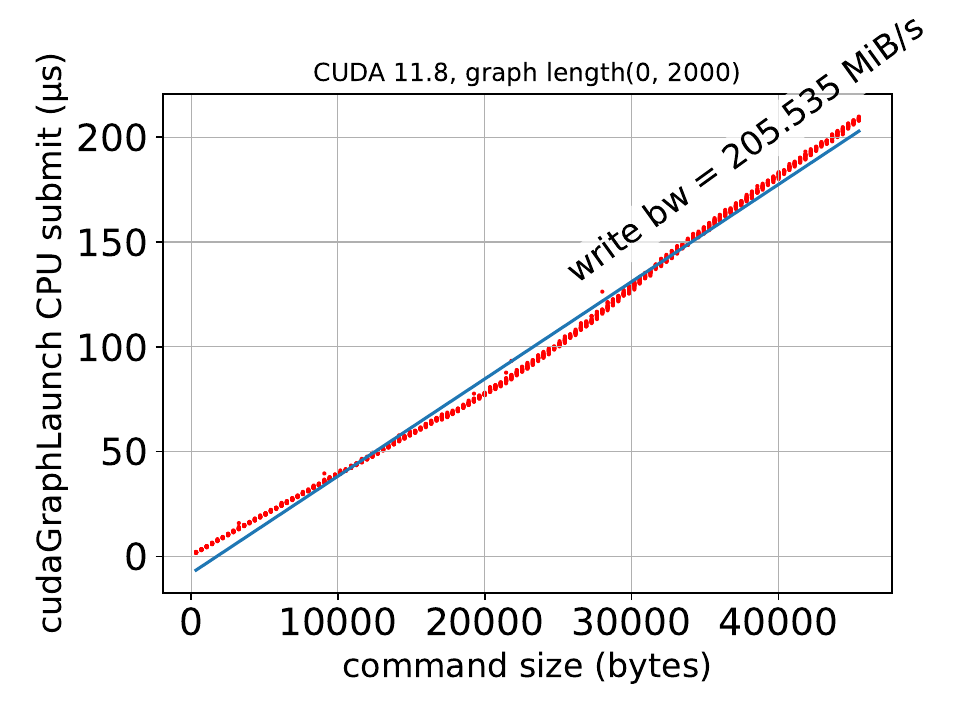}
  \caption{CUDA 11.8, graph length 0–2000}
  \label{fig:11_8_pb_vs_time_0_2000}
\end{subfigure}\hfill
\begin{subfigure}{0.50\columnwidth}
  \centering
  \includegraphics[width=\linewidth]{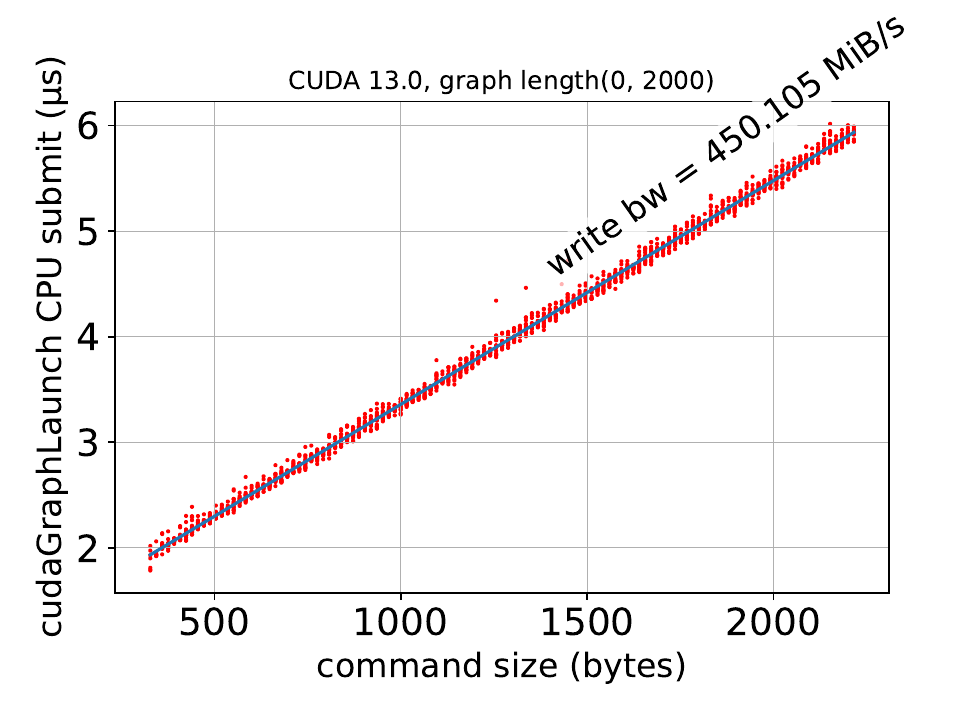}
  \caption{CUDA 13.0, graph length 0–2000}
  \label{fig:13_0_pb_vs_time_0_2000}
\end{subfigure}

\caption{Relationship between pushbuffer command size and CPU submission time for CUDA Graph launches. The annotated slope is a linear fit as an effective write bandwidth (MiB/s). Results are shown for CUDA 11.8 and CUDA 13.0 at two graph-length ranges (0–200 and 0–2000 kernels).}
\label{fig:cugraphmemfitting}
\end{figure}
In \Cref{fig:submission_vs_all}, we present three submission indicators as graph size increases: CPU launch time, total size of commands issued, and the number of doorbell writes. Since a doorbell write is the CPU-side MMIO notification that new GPFIFO entries have been posted to the GPU, we interpret the doorbell-write count as a proxy for the number of distinct command submission cycles performed by the driver. 

Under CUDA~13.0, graph launch time shows only a slight increase, from \SI{1.9}{\micro\second} to \SI{5.9}{\micro\second}, as graph length grows from 1 to 2000. In contrast, CUDA~11.8 starts from a similar startup (\SI{1.8}{\micro\second}) but increases steadily to \SI{209}{\micro\second} at graph length 2000, exhibiting a clear linear growth with respect to graph length. This contrast is mirrored in the total command size: CUDA~11.8 increases from \SI{328}{\byte} to \SI{45476}{\byte}, whereas CUDA~13.0 increases only from \SI{340}{\byte} to \SI{2216}{\byte} over the same range, following the same scaling trend observed in launch time.

It is noticable that when zooming into the short-chain range (0--200) in \Cref{fig:chainlen_vs_doorbells_0_200,fig:chainlen_vs_time_0_200,fig:chainlen_vs_size_0_200}, the growth of launch time and command size not only matches in magnitude, but also shows the same stepwise pattern, most clearly under CUDA~11.8. In \Cref{fig:chainlen_vs_size_0_200}, the command size changes in discrete steps: it remains unchanged over multiple consecutive graph lengths and then jumps at specific breakpoints, yielding a staircase shape of the plot. The launch time in \Cref{fig:chainlen_vs_time_0_200} also shows a similar, approximate staircase behavior, with short plateaus across ranges of graph lengths and intermittent increases.

\begin{figure}[htbp]
\centering
\begin{subfigure}{0.48\columnwidth}
  \centering
  \includegraphics[width=\linewidth]{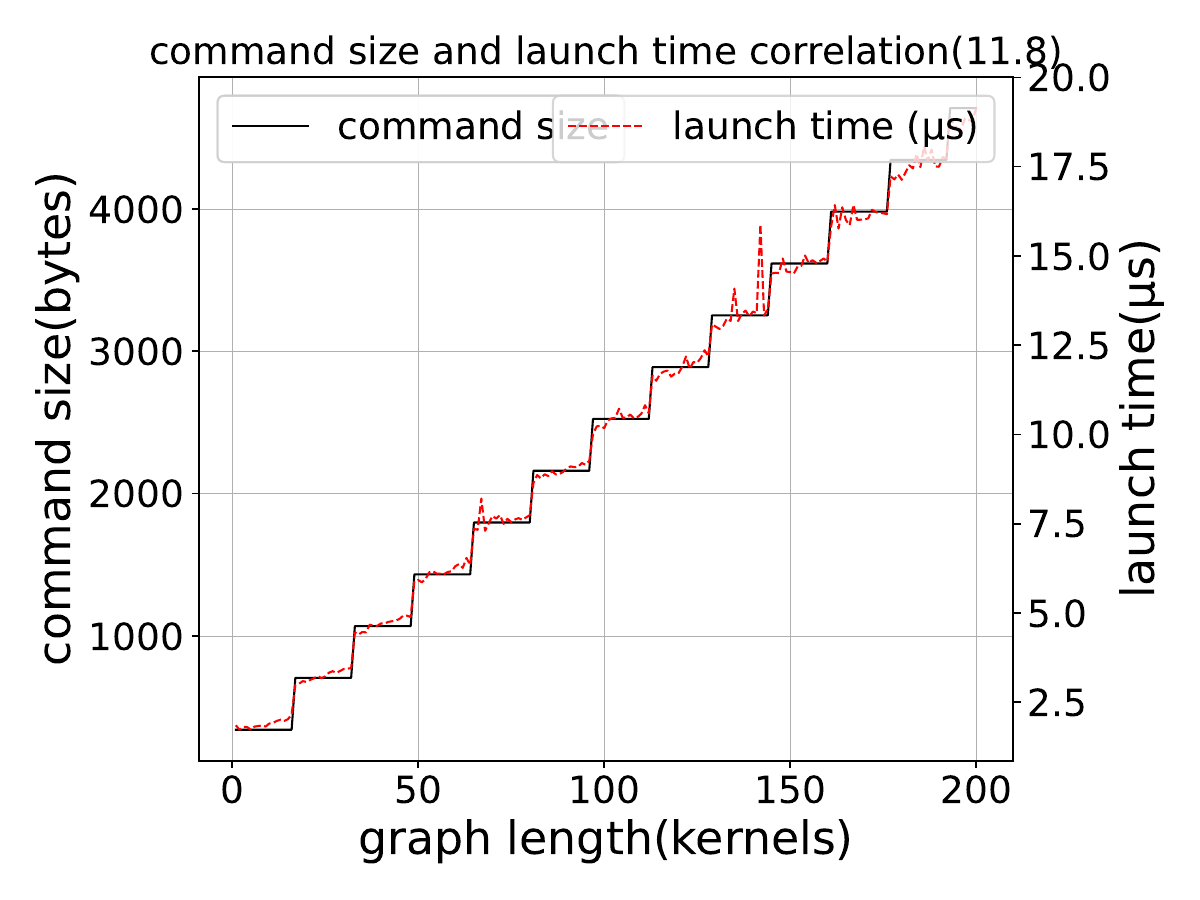}
  \caption{CUDA 11.8}
  \label{fig:pb_time_correlation_0_200_11_8}
\end{subfigure}\hfill
\begin{subfigure}{0.48\columnwidth}
  \centering
  \includegraphics[width=\linewidth]{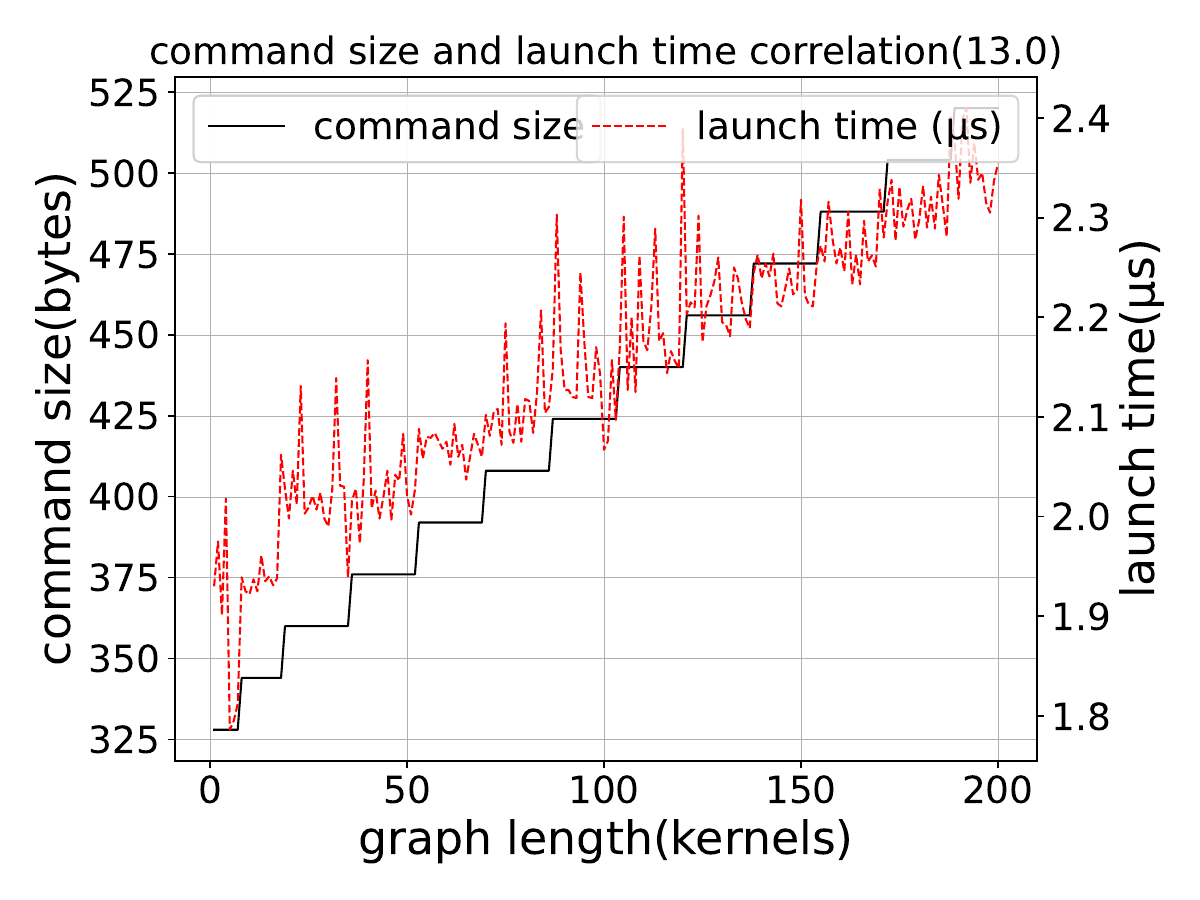}
  \caption{CUDA 13.0}
  \label{fig:pb_time_correlation_0_200_13_0}
\end{subfigure}

\caption{Command size and CPU launch time correlation as a function of CUDA Graph chain length (0–200 kernels), comparing CUDA 11.8 and CUDA 13.0. Left y-axis shows reconstructed pushbuffer command size (bytes) and right y-axis shows measured CPU launch time (µs).}
\label{fig:pb_time_correlation}
\end{figure}

We plot \Cref{fig:pb_time_correlation} to better compare the relationship between reconstructed command size and graph launch time in the short-chain range (1--200) for CUDA~11.8 and CUDA~13.0. Under CUDA~11.8, the step transitions in launch time and command size are closely aligned in graph length. For CUDA~13.0, a similar staircase in launch time is not apparent in this zoomed view: the increase across graph length is modest (192~bytes and \SI{0.5}{\micro\second}), so system jitter becomes comparable to the signal at this time scale. Nevertheless, the overall increasing trend with graph length remains visible for CUDA~13.0 even within this small range.

Given the observations above, the tight coupling between command size and launch time suggests that \texttt{cudaGraphLaunch} overhead is sensitive to the command footprint. From a driver perspective, a larger command stream typically incurs more host-side work to assemble and serialize the submission buffer (e.g mapping the method address to different types of pushbuffer headers), and it also increases the volume of data written through the submission path. While other factors may influence driver--GPU interaction, our CUDA~11.8 vs.\ CUDA~13.0 comparison shows that a smaller command footprint is consistent with substantially lower launch overhead, with agreement not only in overall magnitude but also in the observed fine-grained stepwise behavior. Taken together, the command size could serve as a useful precursor for reasoning about CUDA runtime launch overhead.

\subsubsection{Submission memory access pattern}

Another distinction in command issuing between CUDA~11.8 and CUDA~13.0 is the number of doorbell writes, shown in \Cref{fig:chainlen_vs_doorbells_0_200,fig:chainlen_vs_doorbells_0_2000}. Under CUDA~11.8, as graph length increases, the number of submission doorbell writes during the \texttt{cudaGraphLaunch} API call also increases. This suggests that the driver performs multiple submission cycles, 
In contrast, CUDA~13.0 consistently issues a single doorbell write, indicating a single submission cycle.

Note that in our environment the pushbuffer and GPFIFO reside in different memory domains (\findingref{finding:locality}): host RAM and GPU video memory, respectively. Consequently, the CUDA~11.8 pattern repeatedly alternates the CPU write destination between host memory (pushbuffer construction) and remote GPU memory (GPFIFO/doorbell), as illustrated in \Cref{fig:pattern}, which incurs PCIe TLP (Transport Layer Packet) writes. Under CUDA~13.0, the driver writes the pushbuffer and then notifies the GPU once, resulting in only one transition to remote GPU writes.

To evaluate the efficiency of these two command submission patterns, we compare command size and the resulting graph-launch overhead in \Cref{fig:cugraphmemfitting} across both driver stacks and at different scales. We use a least-squares linear fit and report the fitted slope as an effective write bandwidth (MiB/s) as an indicator of submission efficiency. 

The fitted command-emission bandwidth remains relatively stable across the evaluated command-size ranges for both stacks. For graph lengths 1--200 and 1--2000, CUDA~11.8 achieves 243.97 MiB/s and 205MiB/s, respectively, while CUDA~13.0 achieves 432.16MiB/s and 450.11 MiB/s. Overall, CUDA~13.0 sustains roughly twice the effective bandwidth of CUDA~11.8, showing the impact of different memory writing patterns: swinging between RAM and TLP frequently will reduce the benefit of batched writing and could introduce additional PCIe ordering in the submission path, which can increase host-side launch overhead. 


\section{Conclusion and Future work}

In this work, we study NVIDIA’s GPU command-submission path and break down its cost into hardware and software components. We make this path concrete by identifying where command buffers reside (pushbuffer/GPFIFO locality) and by quantifying the command footprint of each submission. With both the command-delivery path and its acknowledgment path made explicit, driver overhead can be broken down more precisely into stages such as host-memory writes, host--device DMA transfers, host--device MMIO writes, and completion signaling. This in turn raises an important future design question: where command buffers should be placed, whether in device memory, host memory, or a hybrid arrangement as observed in our experiments. Such design trade-offs have received little attention in the literature, despite their likely importance on platforms with different host--device memory characteristics.

With the stage-level view of command submission, it becomes possible to investigate specific driver mechanisms more directly. For data transfer, unlike open-source stacks such as Open MPI, where protocol thresholds are exposed and tunable, the corresponding logic in CUDA has remained opaque and not user-controllable. By directly programming the DMA engine, our method makes the transfer protocol explicit, creating opportunities for tuning and for informing the design of future networked or disaggregated GPU systems. If the measured raw bandwidth over the interconnect is already below the hardware baseline, the bottleneck likely lies in the link itself. If raw DMA-engine performance remains intact but command travel time becomes longer, the issue is more likely the command path and submission pattern. Without this clarity, both cases would appear simply as non-optimal performance, making them hard to reason about and improve.

The same stage-level breakdown also provides useful lessons for reducing GPU launch overhead. As the CUDA Graph case study shows, newer CUDA releases reduce host-side launch cost by reshaping the command stream and submission pattern, thereby reducing CPU involvement in the critical path. This highlights that launch latency can be mitigated through driver-side organization of GPU work.

\bibliographystyle{ACM-Reference-Format}
\bibliography{sample-base}


\end{document}